\documentclass[aps,pra,eqsecnum,noshowpacs,amssymb,nofootinbib]{revtex4}
\usepackage{amsfonts}
\usepackage{graphicx}
\usepackage{bm,amssymb}
\usepackage{epigraph}
\usepackage{amsmath,fontenc}
\usepackage{array}
\usepackage{color}

\graphicspath{{/media/alex/Data/PHYSICS/Figures/}}
\newcommand{\stirling}{\genfrac\{\}{0pt}{}}
\newcommand{\uom}{\underline{\omega}}
\newcommand{\unu}{\underline{\nu}}
\newcommand{\uGa}{\underline{\Gamma}}
\newcommand{\tz}{\tilde{z}}
\newcommand{\td}{\tilde{d}}

\begin{document}	
\title{Theoretical ground for precursors-based molecular spectroscopy}
\author{Alexander Makhlin$^1$}
\author{Panagiotis Papoulias$^2$}
\author{Eugene Surdutovich$^3$}
\email[Corresponding author: ]{surdutov@oakland.edu}
\affiliation{$^{1}$ Rapid Research Inc, Southfield, Michigan 48076, USA}
\affiliation{$^{2}$ Science Seals, LLC, Ann Arbor, Michigan 48105, USA}
\affiliation{$^{3}$ Department of Physics, Oakland University, Rochester, Michigan 48309, USA}
\date{\today}
\begin{abstract}
A theory for excitation of molecular resonances by a train of precursors is developed. Right at the vacuum-medium interface, a train of incident square waves interacts with light electrons and is converted into a train of precursors, which further excite molecular dipoles. Analytic calculations indicate that these excited dipoles generate radiation, including secondary precursors propagating in the backward direction. Encoded in this radiation are proper frequencies of excited molecular dipoles allowing for spectroscopic measurements.
The frequency of the train of incident square pulses can be by several orders of magnitude smaller than the proper frequencies of molecular resonances.
\end{abstract}

\maketitle

\section{ Introduction }\label{sec:Sec1}
	
The notion of precursors (the name adopted from seismology) as a physical entity was introduced in optics by A. Sommerfeld as the solution to an apparent paradox: in the domain of anomalous dispersion the group velocity can exceed the speed of light, $c$, in vacuum. This was obviously in conflict with the special theory of relativity~\cite{First1}. Considering a semi-infinite sinusoidal signal at the interface between vacuum and medium, Sommerfeld proved that the leading part of the signal following the front propagates in the medium with speed $c$. This leading part is known as a precursor. The physics of this phenomenon was attributed to the dynamic nature of the refraction index, $n(\omega)$. The latter cannot differ from unity until the electronic polarization is engaged in response to electromagnetic wave. In recent years, precursors have attracted attention of both experimentalists and theorists. An extensive review is given in books by K. Oughstun~\cite{BS3}. In this paper we propose to use them for the purpose of spectroscopy.

Traditionally, spectroscopic measurements are conducted in continuous mode and assume the availability of  quasi-monochromatic sources of radiation. An underlying assumption is that any properties of measured signals are encoded in the dispersion law of the index of refraction and rely on the availability of high resolution spectral devices, which may not always be the case, e.g. in millimeter range radiation. In this paper, we propose another approach. Reacting to steep wavefronts of the incident electromagnetic field, the medium generates, right at the vacuum-medium interface, short pulses, precursors, with their leading fronts traveling through a medium at the speed of light. Precursors are insensitive to any properties of the medium, except for the ubiquitous electronic polarization.  However, a long train of \textquotedblleft primary precursors\textquotedblright can induce and substantially amplify oscillations in molecular dipoles, which subsequently radiate not only in the forward, but also in the backward direction with respect to the incident signal.

The current study is founded on the theoretical work \cite{precursor1969} of 1969  by G.V.~Skrotsky and his group. Impetus for their work was provided by advances in the generation of ultrashort optical pulses with steep wavefronts and the possibility of measuring of time intervals down to the order of $10^{-14}$~sec \footnote{A successful direct measurement of precursors in a region of anomalous  dispersion was reported only in 2006 \cite{Direct} by a group from Duke University.}. Paper~\cite{precursor1969} studied the formation of a precursor during a traversal of a vacuum-medium interface by the front of a light pulse and its {\em passage through a slab} of matter.
It was found that precursors can be completely separated from an initial semi-infinite harmonic signal and that, sufficiently close to the leading front, an \textquotedblleft  instantaneous frequency\textquotedblright of the precursor's electric field  increases with the thickness of the slab, thus making them less and less sensitive to the properties of a medium. Exactly at the leading front, the amplitude of the electromagnetic field  remains the same at any distance of its propagation regardless of the number of slabs it crosses. In the current study, we build on these physically important facts and suggest that precursors may be utilized
in spectroscopic studies of molecules or detection of various chemical substances.

The approach taken in this study is prompted by a large difference of time scales involved in the procedure of measurement and can be briefly described as follows. Let a train of square pulses with sharp wavefronts be incident on a vacuum-medium interface. Light electrons are immediately accelerated and radiate even before they acquire velocity and displacement. The electronic component of electric polarization at a time immediately following the wavefront can be adequately described by the  ``plasma'' refraction index, $n_e(\omega)$. The scale of this process is determined by the Langmuir frequency $\Omega_e\sim 10^{15} - 10^{16} $ rad/sec corresponding to the density of {\em all} electrons. The electric field of precursors produces an external force in the mechanical equations of motion of elastic molecular dipoles; these equations can be solved exactly. The scale of this process is set by the proper frequency $\omega_0\sim 10^{12}$ rad/sec  and the width $\Gamma_0\ll\omega_0 $ of a particular molecular resonance. The acceleration of the dipole's constituent charges results in a detectable radiation. The field of this radiation is a sum of slowly varying (with the proper frequency of the elastic dipole's oscillation) electromagnetic  fields and of highly oscillating (with the electronic Langmuir frequency) fields of precursors. The proper frequencies of molecular oscillations can be identified by positions of maxima in intensity of backward radiation as functions of duration $T$ of incident pulses (or the frequency $\nu_0\sim 10^{8}-10^{10}$ Hz of the pulses' repetition in the incident train).

The paper is arranged as follows. In Sec.\ref{sec:Sec2} we consider the first and fastest process of formation of primary precursors at the vacuum-medium interface. We begin with the simplest case of a single step and introduce mathematical methods used throughout the paper. We derive an explicit expression for the electric field of a single precursor and trace its evolution in the course of its propagation inside the medium. Then, we consider its passage through an interface and reflection of the incident signal in the form of a single square pulse, which, having both leading and rear fronts, produces two precursors. Finally, we examine propagation of a pulse through a slab of  matter  with finite thickness.

In Sec.\ref{sec:Sec3} and Appendix \ref{app:appA} we  solve the equations of motion for elastically bound charges in the field of a {\em train of primary precursors}, which originate from  an incident train of square pulses. We find an explicit time dependence of the electric dipole moment of the molecule, as well as the {\em ladder of amplitudes} of harmonic oscillations that are induced and amplified by the train of precursors. Oscillating dipoles must radiate.
Their radiation propagates inside a dispersive medium and, eventually, escapes into the vacuum. In Sec. \ref{sec:Sec4} we find the Green's functions that solve the problem of radiation and also explicitly account for the boundary conditions at the interfaces between the medium and vacuum. We find that dipoles radiate both in the forward and backward directions with respect to the direction of propagation of  the trains of incident pulses and of the primary precursors.

The expression for the electric field of the molecular dipoles'  radiation is derived in Sec.\ref{sec:Sec5}. The electric dipole polarization induced by the primary precursors includes two distinct components. One of them is proportional to the field of the entire train of primary precursors, which does not lead to radiation and is a strict analytic result. Its presence can be accounted for by small corrections to the purely electronic refraction index. The second component, also found analytically, is  due to abrupt jumps in the amplitude of the elastic dipoles' oscillations. It bears an anticipated harmonic pattern in addition to yet another train of {\em secondary precursors} radiated in the backward direction.

In Sec.~\ref{subsec:Sec6A} we analyze and interpret the results obtained in Sec.~\ref{sec:Sec5}. A general discussion and outlook follow in Sec.~\ref{subsec:Sec6B}. Appendices~A, B and C present some details of analytical calculations in Secs.~\ref{sec:Sec3} and~\ref{sec:Sec5}. In Appendix~D, a method allowing for numerical calculations elucidating the analytical results obtained in Sec.~\ref{sec:Sec5} and presented in Sec.~\ref{subsec:Sec6A} is shown and discussed.

\section{Formation and propagation of precursors}\label{sec:Sec2}	

In this section, we closely follow Ref.~\cite{precursor1969} gradually changing the setup of the problem. We start with a semi-infinite incident step pulse propagating from vacuum into a medium, then continue with a single rectangular incident pulse. For the rest of the paper we consider a long train of incident square pulses with alternating polarity.  After any wavefront crosses an interface, a purely electronic polarization transforms a signal into a precursor.

For a front of a semi-infinite wave incident on a plane interface between the vacuum and a medium, a steady state of propagation  is reached after some time has elapsed. The electromagnetic field of a steady state satisfies the extinction theorem of Ewald and Oseen \cite{Born,Rosenfeld};
two waves are formed in the medium, a refracted wave with a phase velocity of $c/n$ and a not refracted wave propagating with the speed of light in vacuum. The latter wave exactly cancels out the incident wave in the medium and only a refracted wave is observed. However, a time interval, longer than the characteristic time inherent to the medium, is required for the steady state to form.
During this interval immediately following the wavefront (before the refracted and non-refracted waves are formed), a precursor propagating with the speed of light in the direction of the incident wave is produced.

Traditionally, an electromagnetic signal in a medium is represented by a sum of harmonics. Each harmonic is a stationary signal, which \textquotedblleft knows nothing\textquotedblright of its origin from a limited wave train, and
behaves as a plane wave in a dispersive medium. Its propagation is described by the stationary index of refraction and stationary boundary conditions, as given by the Fresnel formulas. The electromagnetic characteristics of the medium are determined by natural frequencies $\omega_q$ of bound electrons and their relaxation times, $\tau_{rel}\sim 1/\Gamma_0$. Within a time interval of about $2\pi/\omega_q$ from the instant of arrival of the wave front at a given point, excitation and relaxation processes play only a secondary role. From the point of view of the damped classical oscillator model, electrons do not have time to acquire either velocity or displacement with respect to their equilibrium positions. 	

\subsection{Introductory calculations, an incident step signal}\label{subsec:Sec2A}	

We examine the properties of precursors and consider the propagation of various signals in the simplest case, i.e., when a medium has no molecular resonances, while polarization due to light electrons completely determines the index of refraction, $n_e(\omega)$,
\begin{eqnarray}\label{eq:E2.1}
n_e^2(\omega)=1-{\Omega_e^2\over\omega^2},~~ \Omega_e^2={4\pi N_e e^2\over m_e}~,
\end{eqnarray}
where $\Omega_e$ is the Langmuir (plasma) frequency.   Even though in anticipated experiments we expect the incident signal to be a long sequence of alternating square pulses it is instructive to start with
 a single step of unit amplitude, which has a well-known spectral representation,
	\begin{eqnarray}\label{eq:E2.2}
	E_{0}(t,z)=\theta(t-z/c)={-1\over2\pi i}\int_{ia-\infty}^{ia+\infty}
	{d\omega\over \omega}	e^{-i\omega(t-z/c)},~~~~
	\end{eqnarray}
where $\omega/c=k_0$ is the wave vector of propagation in free space.
After the leading wavefront crosses the vacuum-medium interface, the amplitudes of Fourier-components of a signal acquire the transmission factor  ${\mathfrak T}(n_e)$, while the wave vector $k_0=\omega/c$ changes for $k(\omega)=\omega n_e(\omega)/c$.
The electric field of such an incident pulse inside the medium is
\begin{eqnarray}\label{eq:E2.3}
	E'_t(t,z)={-1\over 2\pi i} \int_{ia-\infty}^{ia+\infty}{d\uom\over\uom}{\mathfrak T}[n_e(\omega)] e^{-i\Omega_e[\uom t-\sqrt{\uom^2-1}~\tz] }\propto \theta(t-z/c)~,
\end{eqnarray}
where $\omega n_e(\omega) =\Omega_e\sqrt{\uom^2-1}$, with $\uom=\omega/\Omega_e$ and $\tz=z/c$. The Fresnel coefficients of transmission, ${\mathfrak T}$, and reflection, ${\mathfrak R}$, for partial monochromatic waves (on a plane boundary between medium and vacuum and normal incidence) are well-known~\cite{Born},
\begin{eqnarray}\label{eq:E2.4}
	{\mathfrak T}(n_e)\!=\!{2\over 1+n_e}\!=\!{2\uom\over\uom+\sqrt{\uom^2-1}}={1\over n_e}{\mathfrak T}({1\over n_e}),~~~~{\mathfrak R}(n_e)={1-n_e \over 1+n_e}\!=\!{\uom- \sqrt{\uom^2-1}\over \uom+\sqrt{\uom^2-1}}=-{\mathfrak R}({1\over n_e})~.
\end{eqnarray}

\begin{figure}[h]
	\includegraphics[width=0.45\textwidth]{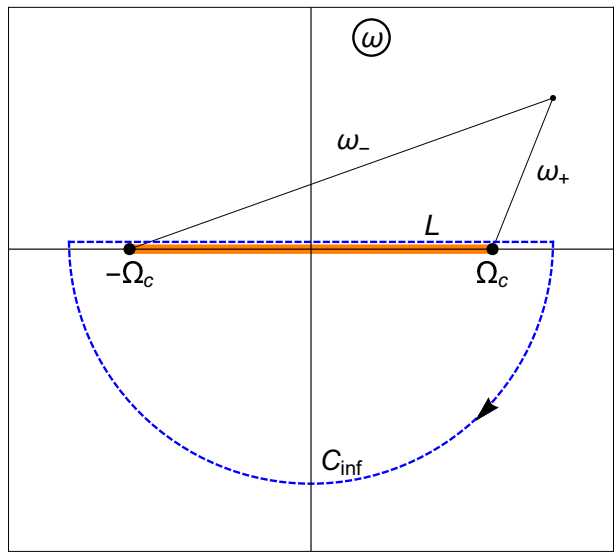}
	\caption{ Cut in $\omega$-complex plane and the integration contour $C_\omega$. }
	\label{Fig:fig1}
\end{figure}

In the integrals like (\ref{eq:E2.3}) the path $L$ of integration along the real axis of $\uom$ can be augmented with a semicircle having an infinite radius, $C_{inf}$, in the lower half-plane of $\uom$, thus forming a closed clockwise contour $C_\omega$ (since $t-z/c>0$, the integral over $C_{inf}$ is zero). The integrand has two branching points at $\omega=\pm\Omega_e$ ($\uom=\pm 1$) and  is double-valued. It will become single-valued after we cut the  complex $\omega$-plane along the segment of the real axis between the branching points.   Since there are no other singularities, one can take for $C_\omega$ any closed path encapsulating the cut (see. Fig.\ref{Fig:fig1}).

In order to compute this contour integral, we resort to the method originally proposed by N.G. Denisov~\cite{Denisov} and used in Ref.~\cite{precursor1969}~\footnote{\label{ftnt:fn2} An indisputable advantage of this method is that in many cases it yields analytic solutions valid throughout all range of time $t$ and distance $z$. Contrary to more popular asymptotic methods of saddle point or steepest descent~\cite{First1,First2}, which provide reasonable approximations only at large times and/or distances, the Denisov's approach works even at the earliest moments of a transient process (which has been reiterated and emphasized in somewhat different context in Ref.\cite{waveguide}).   A theoretical analysis along the traditional guidelines of Sommerfeld and Brillouin (asymptotic calculation of the spectral integrals) has been revisited more than once Ref. \cite{BS1,BS2,BS3}, where the reader can also find an extensive critical review of many other papers.}. A new variable, $\zeta=\uom-\sqrt{\uom^2-1}$, corresponds to that branch of the conformal mapping, $\uom=(\zeta+1/\zeta)/2$, that maps the complex plane of $\uom$ with the cut between branching points $\uom=\pm 1$ onto an {\it exterior of a unit circle} $|\zeta|=1$ in the plane of complex $\zeta$.  The integration contour in the $\zeta$-plane is a circle with the center at its origin; in all cases considered below, it does not enclose any singularities  and it is traversed in the counterclockwise direction. The upper and lower banks of the cut in the $\uom$-plane are mapped onto the upper  and lower semicircle in the $\zeta$-plane, respectively.  The phases of complex functions $\omega_+=\omega-\Omega$ and $\omega_- =\omega+\Omega$ are fixed in such a way that for real $\omega>\Omega_e$, we have ${\rm arg}(\omega_+)={\rm arg}(\omega_-)=0$. Then, for $|\omega|<\Omega_e$, we have ${\rm arg}(\omega_+)=+\pi$ and ${\rm arg}(\omega_-)=0$ on the upper bank of the cut, with ${\rm Re}k_z=0$ and  ${\rm Im}k_z>0$, as expected.

\begin{figure}[h]
		\includegraphics[width=0.9\textwidth]{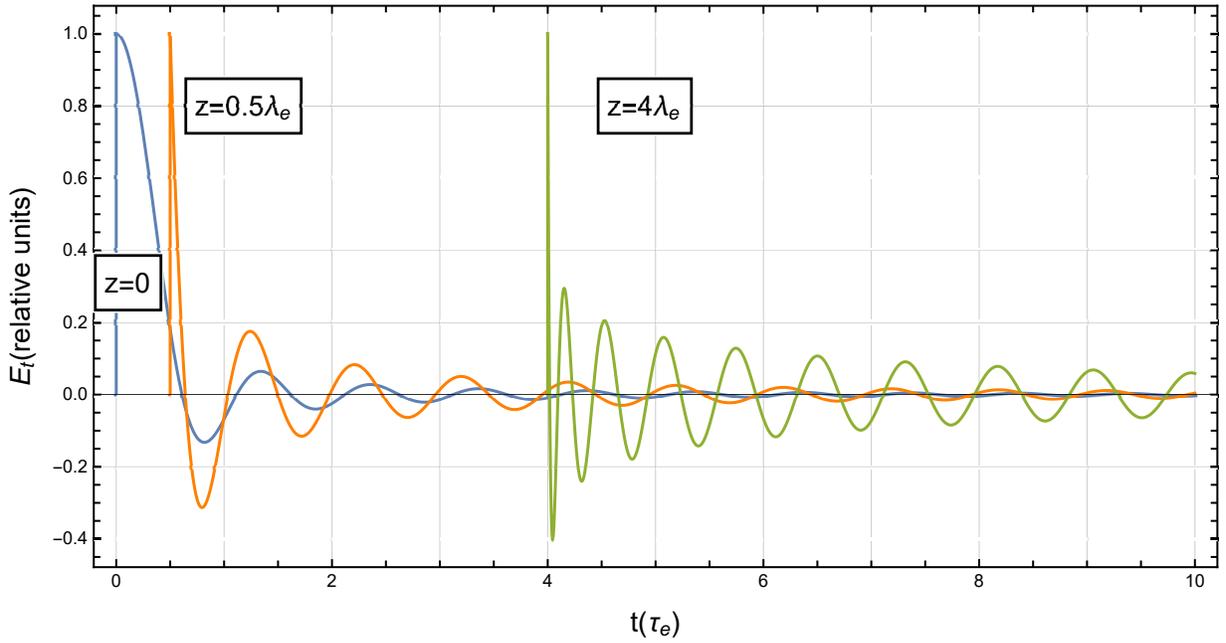}		
		\caption{Plots illustrating the time dependency of precursors formed following a stepwise signal incident on the surface at $z=0$. Time is measured in periods of plasma oscillations ($\tau_e=2\pi/\Omega_e$). Time dependencies are shown for different depths $z$ (in units of $ \lambda_e$).
	}
			\label{Fig:fig2}	
	\end{figure}

It is straightforward to check the following formulae, which will often be used throughout the paper,
\begin{eqnarray}\label{eq:E2.5}
	\uom={1\over 2}({1\over \zeta}+\zeta),~~~\uom n_e(\omega)=\sqrt{\uom^2-1}
	={1\over 2}({1\over \zeta} -\zeta),~~~{d\uom\over\uom}=-{d\zeta\over\zeta} {1-\zeta^2\over 1+\zeta^2} ,~~~~ {d\uom\over\uom}\mathfrak{T}[n_e(\omega)]=-{d\zeta\over\zeta} (1-\zeta^2).
\end{eqnarray}
The phase factor, $e^{-i\Omega_e[\uom t -\sqrt{\uom^2-1}~\tz] }$, in the integrand of (\ref{eq:E2.3}) becomes $e^{-i(\Omega_e/2)[(t-\tz)/\zeta+(t+\tz)\zeta] }$ and the integral now reads as,
\begin{eqnarray}\label{eq:E2.6}
	E'_t(t,z)\!\!=\!\!{\theta(t-\tz)\over 2\pi i}\oint^{(0_+)}{d\zeta\over\zeta}(1-\zeta^2)\exp\bigg{\{-i{\Omega_e\tau\over 2}[{\xi\over\zeta}\!+\!{\zeta\over\xi}]\bigg\}}\!\!=\!\!{\theta(t-\tz)\over 2\pi i}\oint^{(0_+)}{d\zeta\over\zeta}(1-\xi^2\zeta^2)\exp{\bigg\{\!\!-i{\Omega_e\tau\over 2}\big[{1\over\zeta}\!+\!\zeta\big]\bigg\}},~~
\end{eqnarray}
where  $\tau^2=t^2-\tz^2$,  $\xi^2=(t-\tz)/(t+\tz)$.  The factor $\exp{\{-iq( \zeta+1/\zeta )/2\} }$ in the integrand of (\ref{eq:E2.6}) is the generating function for the Bessel functions of an integer order \footnote{\label{ftnt:fn3}This representation differs from the originally referred to by Denisov \cite{Denisov} (and most often used in the literature, e.g. \cite{Watson}, \S2.2 (4)), $$J_n(q)={1\over 2\pi i}\oint^{(0_+)} {dp\over p^{1+n}}e^{(q/2)[p-1/p]}=(-1)^nJ_{-n}(q),$$ by a trivial change of the variable $\zeta=ip$. }
\begin{eqnarray}\label{eq:E2.7}
	{1\over 2\pi i}\oint^{(0_+)}{d\zeta\over\zeta^{1+n}}e^{-i(q/2)[\zeta+1/\zeta]}=(-i)^nJ_n(q)=(+i)^nJ_{-n}(q).
\end{eqnarray}
The exact analytic answer reads,
\begin{eqnarray}\label{eq:E2.8}
	E_t(t,z)\equiv E'_t(t,z)=\theta(t-\tz)[J_0(\Omega_e\tau)+\xi^2J_2(\Omega_e\tau)].
\end{eqnarray}
The results of calculations for Eq.~(\ref{eq:E2.8}) are presented in Fig.~\ref{Fig:fig2}. They are shown as functions of time for different  depth, $z$, inside a medium.
	Several observations reveal the features of precursors that will be important for the rest of our study.
	First, the deeper the leading front penetrates the medium, the sharper the first maximum is and more rapid the first oscillations are. In other words, in the course of propagation the higher-frequency part of the spectrum of the precursor increases, catching up to the leading front. The Langmuir frequency $\Omega_e$ is dominant on a long tail of the precursor and in its full spectrum (see Ref.\cite{precursor1969}).
Second, regardless of the depth $z$, the amplitude at the leading front, $ct=z$, stays the same and equal to the amplitude of the incident signal. Third, the drop of the amplitude of plasma oscillations with time at $ct>z$ decreases with increasing depth.

	\subsection{Incident single rectangular pulse, reflection  and transmission }\label{subsec:Sec2B}	
	
	Next, we consider several examples of interactions between a  rectangular incident pulse and a medium. Such a pulse is described as the difference 	of two step-functions shifted in time by $T$.  In the spectral representation, the incident  pulse is as follows,
	\begin{eqnarray}\label{eq:E2.9}
	E_{0}(t,z)=\theta(t-z/c)-\theta(t-T-z/c)={-1\over2\pi i}\int_{ia-\infty}^{ia+\infty}
	{1-e^{i\omega T}\over \omega}	e^{-i\omega(t-z/c)}d\omega~.
	\end{eqnarray}
	
	\subsubsection{Passage and reflection of a pulse at the  vacuum-medium interface }
	
\begin{figure}[b]
		\includegraphics[width=0.9\textwidth]{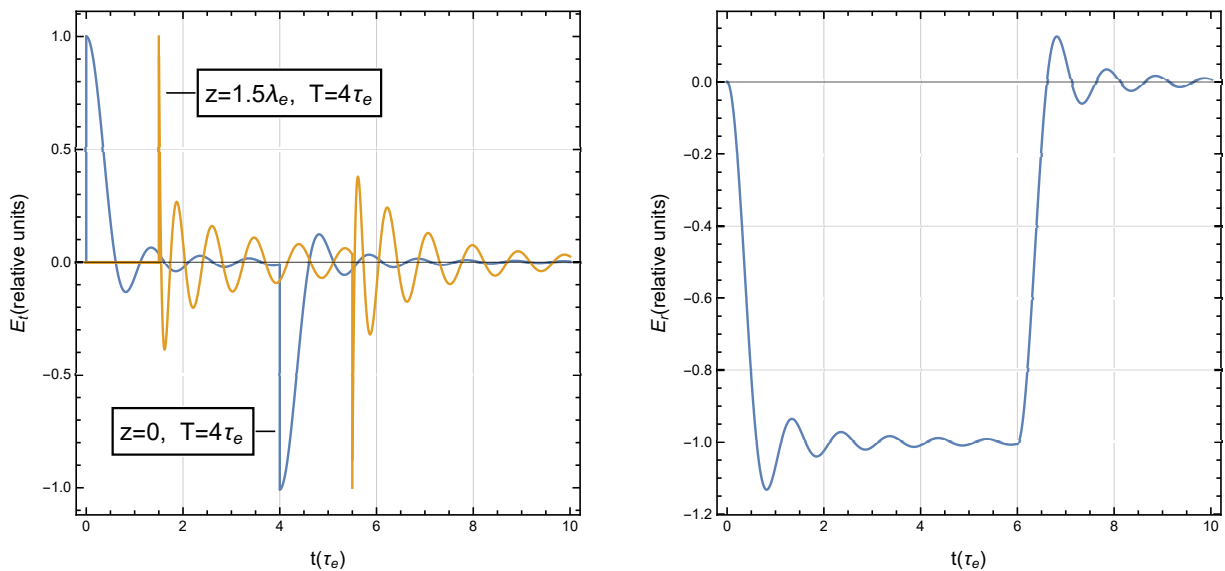}	
		\caption{Two plots illustrating time evolution of  precursors produced by a rectangular pulse. On the left, two fronts of an incident rectangular pulse at two depths $z$. On the right, a  plot of the reflected pulse {\em in the case of normal incident wave.}}
		\label{Fig:fig3}
	\end{figure}

	Substituting in Eq.(\ref{eq:E2.3}) the spectral density (\ref{eq:E2.9}) of a rectangular pulse yields,
	\begin{eqnarray}\label{eq:E2.10}
	E_t(t,z)=E'_t(t,z)-E'_t(t-T,z)={-1\over 2\pi i} \int_{ia-\infty}^{ia+\infty}{d\uom\over\uom}{\mathfrak T}[n_e(\omega)](1-e^{i\omega T}) e^{-i[\omega t-\omega n_e(\omega)z/c] } ~,
	\end{eqnarray}
	where, according to  Eq.~(\ref{eq:E2.8}), $E'_t(t,z)=\theta(t-\tz)[J_0(\Omega_e\tau)+\xi^2J_2(\Omega_e\tau)]$
	and, as previously, $\tau^2=t^2-\tz^2$ and $\xi^2=(t-\tz)/(t+\tz)$.
	This result  is shown in the left panel of Fig.\ref{Fig:fig3} for two different values of $z$, $z=0$ and  $z=1.5\lambda_e$. The leading and the rear fronts of a rectangular pulse generate precursors of the opposite sign.
	The evolution of precursors with depth is similar to that observed in Fig.~\ref{Fig:fig2}.
	
	The spectral form for an electric field of a reflected (back to the vacuum) pulse differs from Eq.~(\ref{eq:E2.3}) by replacement of the transmission coefficient ${\mathfrak T}$  with the reflection coefficient ${\mathfrak R}$ and reversing the direction of propagation, $z\to -z$. Then, for the field $E'_r(t,z)$ reflected at the leading front of the incident pulse,
	\begin{eqnarray}\label{eq:E2.11}
	E'_r(t,z)={1\over 2\pi i}\int_{ia-\infty}^{ia+\infty} {d\uom\over\uom}{\mathfrak R}(n_e)
	e^{-i\Omega_e\uom [t+z/c] }~.
	\end{eqnarray}
	As previously,  we resort to (\ref{eq:E2.5}) to rewrite the integrand in terms  of the variable $\zeta$. Since ${\mathfrak R}(n_e) =(1-n_e)/(1+n_e)=\zeta^2$,  we arrive at the following expression for the reflection of a step-like signal,
	\begin{eqnarray}\label{eq:E2.12}
	E'_r(t,z)={\theta(t+\tz)\over 2\pi i}\oint^{(0_+)}{d\zeta\over\zeta}~{\zeta^2-\zeta^4\over 1+\zeta^2}~\exp{\bigg\{-i{\Omega_e( t+\tz)\over 2}\big[\zeta+{1\over\zeta}\big]\bigg\}} ~~~~~~~~~~~~~~~~~~~~~~~~~~~~~~~~~~~~~~~~\nonumber\\
	= {\theta(t+\tz)\over 2\pi  i}\oint^{(0_+)}{d\zeta\over\zeta}~\sum_{l=0}^\infty(-1)^l
	[\zeta^{2l+2}-\zeta^{2l+4}]~\exp{\bigg\{-i{\Omega_e( t+\tz)\over 2}\big[\zeta +{1\over\zeta}\big]\bigg\}}~~~~~~~~~~~~~~~~~~~~\\
	= -\theta(t+\tz) \sum_{l=0}^\infty \big[ J_{2l+2}(\Omega_e(t+\tz))+J_{2l+4}(\Omega_e(t+\tz))\big]=
	-\theta(t+\tz) [1- 2J_{1}(\Omega_e(t+\tz))/\Omega_e(t+\tz)].\nonumber
	\end{eqnarray}
	Here the last transformation is based on the following identities,  $1=J_0(x)+2J_2(x)+2J_4(x)+...$ and $J_0(x)+J_2(x)=2J_1(x)/x$~\cite{Watson}.
    The exact analytic solution for the reflected field of an incident rectangular pulse is,
	\begin{eqnarray}\label{eq:E2.13}
	E_r(t,z)=E'_r(t,z)-E'_r(t-T,z), ~~~E'_r(t,z)=\theta(t+\tz) [1- 2{J_{1}(\Omega_e(t+\tz))\over\Omega_e(t+\tz)}].
	\end{eqnarray}
	This result is shown in the right panel of Fig.\ref{Fig:fig3}. 	
	An almost static field of a rectangular pulse cannot propagate in a medium with the refraction index (\ref{eq:E2.1}), and is being reflected. The negative sign of the reflected pulse is due to the boundary condition on the interface $z=0$, $E_0+E'_r=E'_t\approx 0$, which is self-evident from visual inspection of the two plots in Fig.\ref{Fig:fig3}.
	
	\subsubsection{Transmission of a pulse through a slab.}
	
	The more realistic problem of passage of a pulse through a slab of thickness $d$ involves two transmission coefficients, one for each interface. For the first interface, as before, the Fresnel coefficient is ${\mathfrak T}(n)$, and ${\mathfrak T}(1/n)$, for the transmission of the pulse from the slab into the vacuum at $z=d$,
	\begin{eqnarray}\label{eq:E2.14}
	E'_d(t,z)={1\over 2\pi i}\int_{ia-\infty}^{ia+\infty}{d\omega\over\omega}{\mathfrak T}(n_e) {\mathfrak T}(1/n_e)  e^{-i\omega t+i\omega(\tz-\td)+i\omega n(\omega)\td },
	\end{eqnarray}
	where $\tz=z/c$ and $\td=d/c$~\footnote{\label{ftnt:fn4}Multiple reflections in the slab are ignored.}. By virtue of Eqs.(\ref{eq:E2.4}) and (\ref{eq:E2.5}), in terms of variable $\zeta$, the product $(d\omega/\omega){\mathfrak T}(n_e) {\mathfrak T}(1/n_e) =4\sqrt{\uom^2-1} (\uom-\sqrt{\uom^2-1})d\uom$ becomes $-(d\zeta/\zeta)(\zeta^2-1)^2$.
	\begin{figure}[t]	
		\includegraphics[width=0.45\textwidth]{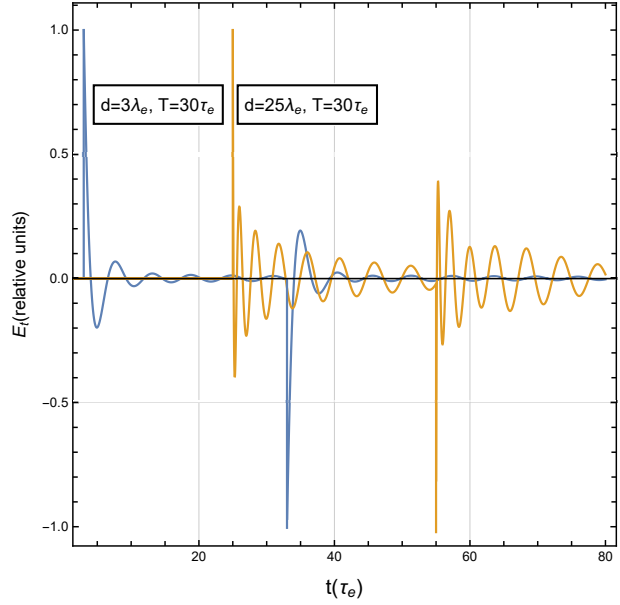}  
		\caption{Plot illustrating the time dependency of precursors that passed through a slab with thickness $d=5\lambda_e$  and $d=25\lambda_e$. The duration of the incident pulse is $T=30\tau_e$.}
		\label{Fig:fig4}
	\end{figure}
	Hence,  the method outlined in Sec.~\ref{subsec:Sec2A} yields,
	\begin{eqnarray}\label{eq:E2.15}
	E'_d(t,z)={\theta(t-\tz)\over 2\pi i}\oint^{(0_+)}{d\zeta\over\zeta}(1-2\xi^2\zeta^2 +\xi^4\zeta^4)\exp{\bigg\{\!\!-i{\Omega_e\tau\over 2}\big[{1\over\zeta}+\zeta\big]\bigg\}} \nonumber\\
	=  \theta(t-\tz)[J_0(\Omega_e\tau)+2\xi^2J_2(\Omega_e\tau)+\xi^4J_4(\Omega_e\tau] ,~~~~~~~~~~~~~~~~~
	\end{eqnarray}
	where $\tau^2=(t-\tz)(t-\tz+2\td)$, $\xi^2=(t-\tz)/(t-\tz+2\td)$, $z\geq d$. For a rectangular pulse $ E_{d}(t,z) =E'_{d}(t,z)-E'_{d}(t-T,z)$, see Fig.\ref{Fig:fig4}. This is precisely the result obtained in Ref.~\cite{precursor1969} under the assumption that the {\em harmonic wave} experiences total internal reflection on the second boundary of the slab.  This can be expected since plasma is optically less dense than vacuum, $n_e(\omega)<1$, and only the precursor passes through.  Also the figure clearly indicates that the thicker the slab is, the sharper are the leading and rear fronts of the  precursors transmitted through a slab into the vacuum.
	
\section{ Excitation of molecular resonances by primary precursors.}\label{sec:Sec3}

In this section we examine the behavior of charges, which form molecular dipoles, in the field of primary precursors. These dipoles become the sources of secondary radiation that carries the desired information about important parameters of dipoles and can be detected.

Let us consider a heavy elastic molecular dipole in the electric field $E_t(t|z_0)$ of a precursor  {\em created at the interface between vacuum and medium by plasma oscillations of light electrons}. Its equation of motion can be written down as follows,
\begin{eqnarray}\label{eq:E3.1}
\ddot{X}(t,z)+2\Gamma_0\dot{X}(t,z)+\omega_m^2X(t,z)=qE_t(t,z)/M~,
\end{eqnarray}
where $M$ and $q$ are effective mass and charge of the dipole, $X$ is its displacement, $\omega_m$ and $\Gamma_0$ are its proper frequency and width.

Let a train of square pulses of duration $T$ be incident perpendicular on the boundary at the point $z=0$ and time $t=0$. Eq.(\ref{eq:E2.2}) can now be generalized as,
\begin{eqnarray}\label{eq:E3.2}
E_0(t,z)={\cal  E}_0[\theta(t-\tz)-2\theta(t-T-\tz)+2\theta(t-2T-\tz)-\dots]
= {-{\cal E}_0\over2\pi i}\int_{-\infty}^{+\infty}{d\omega\over\omega} \sum_{m=1}^{m_p}(-1)^m \epsilon_m e^{im\omega T}e^{-i\omega(t-\tz)}~,~~~
\end{eqnarray}
where $\epsilon_m$ is a so-called Neumann symbol: $\epsilon_m=1$ for $m=0$ and $\epsilon_m=2$ for $m\neq 0$; $m_p=m_p(t)$ is the number of pulses that have passed the boundary $z=0$ by the time $t$. A dipole located at $z_0$ {\it inside the medium} is exposed to the electric field,
\begin{eqnarray}\label{eq:E3.3}
E_t(t,z_0)={\cal E}_0\sum_{m=0}^{m_p}\epsilon_m(-1)^m \theta(t_\ast-mT)E^\prime_t(t_\ast-mT),
\end{eqnarray}
where $t_\ast=t-\tz_0>0$ and, according to Eqs.(\ref{eq:E2.3}) and (\ref{eq:E2.8}.)
\begin{eqnarray}	
E_t^\prime(u)={-{\cal E}_0\over 2\pi i} \oint_{C^-_\omega}{d\uom\over\uom}
{\mathfrak T}[n_e(\omega)]e^{-i\omega u} e^{-i[\omega-\omega n_e(\omega)]\tz_0 } ={\cal E}_0\theta(u)\big[J_0(\Omega_e\sqrt{u(u+2\tz_0)})+{u\over u
	+2\tz_0} J_2(\Omega_e\sqrt{u(u+2\tz_0)})\big].\nonumber
\end{eqnarray}
For a dipole located at a distance $z_0$ from the interface between vacuum and the medium, the general solution of Eq.(\ref{eq:E3.1}) reads as follows,
\begin{eqnarray}\label{eq:E3.4}
X(t|z_0)={q\over M}e^{-\Gamma_0t}\int_{t_0}^te^{\Gamma_0t'} {\sin\omega_0(t-t')\over\omega_0} E_t(t',z_0)dt' +e^{-\Gamma_0t}[b_c(t_0|z_0)\cos\omega_0t+b_s(t_0|z_0)\sin\omega_0t]~,
\end{eqnarray}
where $\omega_0^2=\omega_m^2-\Gamma_0^2$ and  constants $b_c$ and $b_s$ are chosen to satisfy the initial conditions at $t=t_0=z_0/c=\tz_0$.
If this dipole before being exposed to precursors' field was at rest, then $X(\tz_0|z_0)=\dot{X}(\tz_0|z_0)=0$, and, consequently, $b_c=b_s=0$. Equation (\ref{eq:E3.4}) for this dipole becomes
\begin{eqnarray}\label{eq:E3.5}
X(t|z_0)={q\over M}\int_{\tz_0}^t e^{-\Gamma_0(t-t')} {\sin\omega_0(t-t')\over\omega_0}E_t(t',z_0)dt'=	
{q \over M} \int_0^{t_*}e^{-\Gamma_0(t_\ast-t^\prime_\ast)}{\sin\omega_0(t_\ast-t^\prime_\ast)\over\omega_0
}E_t(t'_*,z_0)dt'_*~,
\end{eqnarray}
where $t_\ast=t-\tz_0$ and $t^\prime_\ast=t'-\tz_0$.

The source (\ref{eq:E3.3}) in Eqs.(\ref{eq:E3.1}) and  (\ref{eq:E3.5}) toggles sign abruptly with each passing pulse  and is piecewise continuous.  In order for the general solution (\ref{eq:E3.4}) of Eq.(\ref{eq:E3.1}) to be continuous and differentiable throughout entire time $t$, we first associate the constants $b_c(t_0)$ and $b_s(t_0)$ with $X_{(m_p)}(m_pT)$ and $\dot{X}_{(m_p)}(m_pT)$ (see Eqs.(\ref{eq:A.6})). For the time interval  $m_p T<t_\ast<(m_p+1)T$ we obtain in (\ref{eq:A.7}) a continuous and differentiable function for every $t_\ast$. At the end $(m_p+1)T$ of this time interval (\ref{eq:A.7}) yields a recursion relation connecting $X_{(m_p)}((m_p+1)T)=X_{(m_p+1)}((m_p+1)T)$ and $\dot{X}_{(m_p)}((m_p+1)T)= \dot{X}_{(m_p+1)}((m_p+1)T)$ with $X_{(m_p)}(m_pT)$ and $\dot{X}_{(m_p)}(m_pT)$.

Technical part of the cumbersome calculations for  $X(t|z_0)$ and its first time derivative $\dot{X}(t|z_0)$ is described in Appendix \ref{app:appA}, where we also derive recurrence relations (\ref{eq:A.8})  between the  ${X}_{(m_p)}(m_p|z_0)$ and $\dot{X}_{(m_p)}(m_p|z_0)$ with adjacent numbers $m_p$. In this way we obtain a {\em ladder of amplitudes} $b_c(m_p)$ and $b_s(m_p)$ in Eqs. (\ref{eq:E3.4}) and (\ref{eq:E5.13}) for $m_pT<t<(m_p+1)T$.

Computation of the radiation of molecular dipoles, requires determination of $\ddot X(t|z_0)$ of the constituent charges;  by virtue of (\ref{eq:A.7}),
\begin{eqnarray}\label{eq:E3.6}
\ddot{X}_{(m_p)}(t|z_0) ={{\cal E}_0q\over M} \sum_{m=0}^{m_p(t_*)}\epsilon_m(-1)^m \theta(t_*-mT)E'_t(t_*-mT)
~~~~~~~~~~~~~~~~~~~~~~~~~~~~~~~~~~~~~~~~~~~~~~~~~~~~~~~~~~~~\nonumber \\
- {{\cal E}_0q\over M} \sum_{m=0}^{m_p(t_*)}\epsilon_m(-1)^m \omega_0 \int_{mT}^{t_*} e^{-\Gamma_0(t_*-t')}
\bigg[(1-{\Gamma_0^2\over\omega_0^2})\sin[\omega_0(t_*-t')]+2{\Gamma_0\over\omega_0}
\cos[\omega_0(t_*-t')]\bigg]\theta(t'-mT)E'_t(t'-mT)dt'   \nonumber \\
+\omega_0^2 \bigg\{ -\bigg[(1+{\Gamma_0^2\over\omega_0^2})X_{(m_p)}(m_p T) +2{\Gamma_0\over\omega_0} {\dot{X}_{(m_p)}(m_p T) \over\omega_0}\bigg] \cos\omega_0 (t_*-m_pT) ~~~~~~~~~~~~~~~~~~~~~~~~~~~~~~~~~~~~~~~~~~~~~~\\
+\bigg[{\Gamma_0\over\omega_0}(1+{\Gamma_0^2\over\omega_0^2})X_{(m_p)}(m_p T) -(1-{\Gamma_0^2\over\omega_0^2})  {\dot{X}_{(m_p)}(m_p T)\over\omega_0}\bigg]   \sin\omega_0 (t_*-m_pT)\bigg\}e^{-\Gamma_0(t_*-m_pT)}~. \nonumber
\end{eqnarray}
The second derivative of the {\em density of the dipole polarization} now is $4\pi\ddot{\cal P}_{mol}(t) =4\pi q \langle N_q \ddot{X}(t)\rangle$. We group $\ddot{\cal P}_{mol}(t|z_0)$ into the three terms,
$\ddot{\cal P}_{mol}(t|z_0) = \ddot{\cal P}_a(t|z_0) + \ddot{\cal P}_b(t|z_0)+\ddot{\cal P}_c(t|z_0)$,
\begin{eqnarray}\label{eq:E3.7}
4\pi\ddot{\cal P}_a(t|z_0) = \Omega_q^2{\cal E}_0\sum_{m=0}^{m_p(t_*)}\epsilon_m(-1)^m E'_t(t_*-mT),~~~~~ ~~~~~~~~~~~~~~~~~~~~~~~~~~~~~~~~~~~~~~~~~~~~~~~~~~~~~~~~~~~~~~~{\rm  (a)} \nonumber \\
4\pi\ddot{\cal P}_b(t|z_0) = -  \Omega_q^2{\cal E}_0\sum_{m=0}^{m_p(t_*)}\epsilon_m(-1)^m \omega_0 \int_{mT}^{t_*} e^{-\Gamma_0(t_*-t')}~~~~~~~~~~~~~~~~~~~~~~~~~~~~~~~~~~~~~~~~~~~~~~~~~~~~~~~~~~~~~~~~\\
\times\bigg[(1-{\Gamma_0^2\over\omega_0^2})\sin[\omega_0(t_*-t')]
+2{\Gamma_0\over\omega_0}
\cos[\omega_0(t_*-t')]\bigg]E'_t(t'-mT)dt',~~~~~~~~~~~~~~~~~~~~~~~{\rm  (b)}\nonumber\\
4\pi\ddot{\cal P}_c(t|z_0)=4\pi \Omega_e^2 \uom_0^2 e^{-\Gamma_0(t_*-m_pT)}\bigg\{ C_1(m_pT)\cos\omega_0 (t_*-m_pT)
+C_2(m_pT)  \sin\omega_0 (t_*-m_pT)\bigg\}, ~~~~~~~{\rm  (c)}\nonumber
\end{eqnarray}
where $\Omega_q^2=4\pi q^2 N_q/M$ and
\begin{eqnarray}\label{eq:E3.8}
C_1(m_pT)=- \bigg[(1+{\Gamma_0^2\over\omega_0^2}){\cal P}(m_p T)+2{\Gamma_0\over\omega_0} {\dot{\cal P}(m_p T) \over\omega_0}\bigg],~~
C_2(m_pT)=\bigg[{\Gamma_0\over\omega_0} (1+{\Gamma_0^2\over\omega_0^2}){\cal P}(m_p T) -(1-{\Gamma_0^2\over\omega_0^2})  {\dot{\cal P}(m_p T)\over\omega_0}\bigg],~~~
\end{eqnarray}
For $t_*<T$ (and $m_p=0$) we have ${\cal P}_c(t|z_0)=0$. The difference between these three parts of ${\cal P}_{mol}$ will be discussed in details when we will be looking at their contributions to the field of the dipole's radiation in Sec.\ref{sec:Sec5}.
	
	\section{ Radiation emitted by excited  molecular resonances: General equations}\label{sec:Sec4}	

The goal of this and the following sections is to find an explicit form for the field of radiation caused by the polarization field derived in the previous section.	We consider the radiation due to  uniformly distributed molecular dipoles of number density $N_q$ in an infinitely thin slab of thickness $\Delta z_0$ perpendicular to the $z$-axis. The electric field of their radiation, ${\cal E}_{rad}={\cal E}$ satisfies the wave equation,
\begin{eqnarray}\label{eq:E4.1}
{\partial^2{\cal E}(t,z)\over \partial z^2}-{1\over c^2}{\partial^2{\cal E}(t,z) \over \partial t^2}
={4\pi\over c^2} [\ddot{\cal P}_e(t,z)+\ddot{\cal P}_{mol}(t,z)]
\end{eqnarray}
where ${\cal P}_e(t,z)$ and ${\cal P}_{mol}(t,z)$ are the electronic and molecular components of the electric polarization, respectively. The former,  $\ddot{\cal P}_e(t,z)$, is determined from the equation of motion of free charges,
\begin{eqnarray}
4 \pi\ddot{\cal P}_e(t,z) = 4\pi eN_e\ddot{X}_e(t|z) =4\pi eN_e ~(e/m){\cal E}(t,z)=\Omega_e^2 {\cal E}(t,z). ~~~\nonumber
\end{eqnarray}
The latter, $\ddot{\cal P}_{mol}(t,z)$, was computed in Sec.\ref{sec:Sec3} as the response of molecular dipoles to the field of primary precursors. In the adopted approximation, all effects of the electronic polarization can be  incorporated in the refraction index $n_e(\nu)$, so that ${\cal P}_e(\nu) =\kappa_e(\nu){\cal E}_(\nu)$ and $n_e^2(\nu)=1+4\pi\kappa_e(\nu) =1-\Omega_e^2/\nu^2$. Thus, we are dealing not with the emission of electromagnetic field in vacuum, but rather with the excitation of plasma waves that have well-defined wave fronts and where  electrons are involved in a collective process with the  electric field. The incident pulses excite these waves producing primary precursors at the interface with vacuum. When they reach and excite molecular resonances in the interior of a medium, the latter must radiate. This radiation propagates in a dispersive medium and it must cross an interface where it exits into the vacuum. As will be shown in Sec.\ref{sec:Sec5}, by some of its properties this secondary radiation resembles primary precursors.

Let us assume, for the sake of simplicity,  that molecular dipoles occupy an infinitely thin layer
at depth $z_0$, so that the source surface density in Eq.~(\ref{eq:E4.1}) is $(4\pi/ c^2)\ddot{\cal P}_{mol}(t,z)=(4\pi/ c^2) N_q q \ddot{X}(t|z) \delta(z-z_0)\Delta z_0$.
After applying a Fourier transform with respect to time, equation (\ref{eq:E4.1})  reads as,
\begin{eqnarray}\label{eq:E4.2}
{\partial^2{\cal E}(\nu,z|z_0)\over\partial z^2}+{\nu^2\over c^2}n_e^2(\nu)
{\cal E}(\nu,z|z_0)  ={4\pi\over c^2} \ddot{\cal{P}}_{mol}(\nu,z) \delta(z-z_0) \Delta z_0
\end{eqnarray}
where
\begin{eqnarray}
\ddot{\cal{P}}_{mol}(\nu,z)=\int_{-\infty}^{+\infty}\ddot{\cal{P}}_{mol}(t,z)e^{i\nu t}dt=e^{i\nu\tz_0} \int_{-\infty}^{+\infty} \ddot{\cal{P}}_{mol}(t,z)e^{i\nu t^\ast}dt^\ast. \nonumber
\end{eqnarray}
The solution to the Eq.(\ref{eq:E4.1}) can be found via its Green's function, $G(\tau,z;t,z_0) $,
\begin{eqnarray}\label{eq:E4.3}
{\cal E}_{rad}(\tau,z)=\int G(\tau,z;t,z_0) \cdot\frac{4 \pi}{c^2}\ddot{\cal{P}}_{mol}(t,z_0) dz_0 dt~,
\end{eqnarray}
In order to find its explicit expression, let us perform the Fourier transform of Eq.(\ref{eq:E4.2}) with respect to coordinate $z$. This results in
\begin{eqnarray}\label{eq:E4.4}
-k^2{\cal E}(\nu,k|z_0)+{\nu^2\over c^2}n_e^2(\nu){\cal E}(\nu,k|z_0)={4\pi\over c^2} \ddot{\cal{P}}_{mol}(\nu,z_0) e^{-ikz_0}\Delta z_0~,
\end{eqnarray}
which is an algebraic equation with respect to ${\cal E}(\nu,k|z_0)$. 	Hence, the electric field {\it inside the medium} radiated by molecular dipoles at all depths $z_0$ can be obtained as the double inverse Fourier transform of (\ref{eq:E4.4}), which then can be integrated over all the radiating dipoles,
\begin{eqnarray}\label{eq:E4.5}
{\cal E}(\tau,z) = {-4\pi \over  (2\pi)^2}\int{\sf d}z_0 \int_{-\infty}^{+\infty}d\nu  \int_{-\infty}^{+\infty}dk   {e^{-i[\nu\tau-k(z-z_0)]}\over c^2k^2- \nu^2n_e^2(\nu)} \ddot{\cal{P}}_{mol}(\nu,z_0).
\end{eqnarray}
We start the calculation of this integral with the integration over $k$ along the real $k$-axis that can be reduced to an integral over a closed contour in the complex $k$-plane ($k=k'+ik''$). The choice of a contour depends on the direction of radiation from the layer of dipoles. Indeed, since $e^{ik(z-z_0)}= e^{ik'(z-z_0)}  e^{-k''(z-z_0)}$, for the emission in the forward direction, $z>z_0$, we choose to close the contour of integration in the upper half-plane, where $k''>0$. For the emission backwards, $z<z_0$, the contour should be closed in the lower half-plane. Technically, these requirements can be implemented by specifying the Green function in the $k$-plane as $[c^2k^2-\nu^2 +\Omega_e^2 +i\varepsilon_z]^{-1}$, where $ i\varepsilon_z$ is an infinitesimal imaginary addition to the wave vector $k$ (compare with the well-known causal Feynman's Green's function of QED and also comprehensive analysis of the radiation principle in dispersive medium in Ref.\cite{Bolotovsky}). Then the poles corresponding to the propagation in the forward and backward directions lay slightly below  and above the real axis, respectively. Performing the $k$-integration by the method of residues in these two cases we end up with
\begin{eqnarray}\label{eq:E4.6}
\int_{-\infty}^{+\infty}dk {e^{ik(z-z_0)}\over c^2k^2- \nu^2n_e^2(\nu)} =
{2\pi i\over 2c\nu n_e(\nu)}\big[\theta(z-z_0)e^{i\nu n_e(\nu)(z-z_0)/c}
+\theta(z_0-z)e^{-i\nu n_e(\nu)(z-z_0)/c }  \big]~,
\end{eqnarray}
where the first and the second term in brackets correspond to the emission in the forward and backward directions, respectively.  We are interested in the field outside the medium that occupies the slab $0<z<d$.

To get the field emitted forward, for $z>d$, we must cut off the propagation in the slab at a depth $z=d$, incorporate an additional Fresnel coefficient ${\mathfrak T}[1/n_e(\nu)]$ and continue propagation for the extra distance $z-d$ in free space. To get the field emitted backwards, for $z<0$, we must account for the in-medium propagation for the distance $z_0$, incorporate an additional Fresnel coefficient ${\mathfrak T}[1/n_e(\nu)]$ and continue propagation for the extra distance, $z<0$, in free space. The electric field for either direction reads,
\begin{eqnarray}\label{eq:E4.7}
{\cal E}(\tau,z>d)= - {i\over c}\int_0^d{\sf d}z_0 \int_{-\infty}^{+\infty}{d\nu\over \nu n_e(\nu)} e^{-i\nu\tau}\ddot{\cal{P}}_{mol}(\nu|z_0)  e^{i\nu n_e(\nu)(d-z_0)/c} \mathfrak{T}(1/n_e(\nu)) e^{i\nu(z-d)/c}
,~~~~{\rm (a)}\nonumber\\
{\cal E}(\tau,z<0)= - {i\over c}\int_0^d{\sf d}z_0\int_{-\infty}^{+\infty}{d\nu\over \nu n_e(\nu)} e^{-i\nu\tau}\ddot{\cal{P}}_{mol}(\nu|z_0)e^{i\nu n_e(\nu)\tz_0} \mathfrak{T}(1/n_e(\nu))
e^{-i\nu z/c},~~~~~~~~~~~~~~~{\rm (b)}
\end{eqnarray}
where  ${\mathfrak T}(1/n_e)=n_e{\mathfrak T}(n_e)$ and the integral over real axis in the complex $\nu$-plane can be transformed into an  integral over a clockwise contour $C_\nu^-$ closed by an arc of a large radius in the lower half-plane. The path we took to obtain this result accounts for that fact, that normal modes of our problem are not plane waves in the infinite medium. They satisfy the boundary conditions at the interfaces $z=0$ and $z=d$, which violates translation symmetry in $z$-direction. Furthermore, the radiation of molecular dipoles depends, as does the field of primary precursors, on the depth $z_0$ of a particular dipole.

If we  express $\ddot{\cal P}(\nu|z_0)$ in terms of $\ddot{\cal P}(t|z_0)$, Eqs.~(\ref{eq:E4.7}) acquire the form (\ref{eq:E4.3}), where $G(\tau,z;t,z_0)$ are the corresponding {\it retarded Green's functions} that propagate radiation of the source, $\ddot{\cal{P}}_{mol}(t|z_0)$, i.e., of the dipoles induced by precursors at  ($z_0$, $t$) towards the points of observation ($z$, $\tau$) on either side of a the slab
\begin{eqnarray}\label{eq:E4.8}
G(\tau,z>d;t,z_0)=  -{ i\over 4\pi}\oint_{C_\nu^-}{d\nu\over \nu}  \mathfrak{T}[n_e(\nu)]\cdot  e^{-i\nu[(\tau-t)- (z-d)/c]}e^{i\nu n_e(\nu)(d-z_0)/c},~~~~{\rm (a)}\nonumber\\
G(\tau,z<0;t,z_0)= -{i\over 4\pi}\oint_{C_\nu^-}{d\nu\over \nu} \mathfrak{T}[n_e(\nu)] \cdot  e^{-i\nu(\tau-t-|z|/c)}e^{+i\nu n_e(\nu)\tz_0} .~~~~~~~~~~~~~~~{\rm (b)}
\end{eqnarray}
Equation (\ref{eq:E4.8}) describes  propagation of the radiated electromagnetic field  accounting for the boundary conditions on each interface with the vacuum. These expressions are similar to integrals (\ref{eq:E2.3}) and  (\ref{eq:E2.6}).
They also set up the {\em upper limits} $t_{max}$ of a subsequent integration over $dt$ in Eq.(\ref{eq:E4.3}). These conditions, $\tau>t+|\tz|+\tz_0$ for the emission backward, and  $\tau>t+\tz-\tz_0 $ for the dipole radiation forward, mean that there can be no signal until the leading front of the dipole radiation reaches the point $(\tau, z)$ of an observation. Only the processes in the dipole that took place at $t<\tau-|\tz|-\tz_0$ can affect the detector at time $\tau$. In both cases the path of integration $d\nu$ can be closed by a semicircle in the lower half-plane. The {\em lower limits} $t_{min}(m)=\tz_0+mT$ is the time when the $m$-th pulse hits the dipole. Notably, these Green's functions depend only on the difference $\tau-t$.

Following the scheme of Sec.\ref{sec:Sec3}, one can compute these integrals by mapping the complex plane $\nu$ onto the exterior of a unit circle in the complex plane $\zeta$, so that $\nu=(\Omega_e/2)\big(1/\zeta +\zeta\big)$ (c.f. Eqs.(\ref{eq:E2.5}) ). The result reads as follows,
\begin{eqnarray}\label{eq:E4.9}
G(\tau,z;t,z_0)=   {1\over 2}\cdot{1\over 2\pi i} \oint{d\zeta\over\zeta} (1-\zeta^2)e^{-i{\Omega_e\rho\over 2} [{\mu\over\zeta}+{\zeta\over\mu}]}
={ \theta(\mu\rho)\over 2} [J_0(\Omega_e\rho)+\mu^2J_2(\Omega_e\rho)],
\end{eqnarray}
where $\rho^2=(\tau-t-|\tz|)^2-\tz_0^2$, $\mu^2=(\tau-t-|\tz|-\tz_0)/(\tau-t-|\tz|+\tz_0)$,  $\mu\rho=\tau-t-(|\tz|+\tz_0)$  for $G(\tau,z<0;t,z_0)$ that describes the propagation at the distance $z_0+|z|$ backward, and  $\rho^2=[(\tau-t)- (z-d)/c]^2-(d-z_0)^2/c^2$, $\mu^2=[(\tau-t)- (z-d)/c-(d-z_0)/c]/[(\tau-t)- (z-d)/c+(d-z_0)/c]$, $\mu\rho=\tau-t-(\tz-\tz_0)$ for $G(\tau,z>d;t,z_0)$,  that describes the propagation in forward direction at a distance $z-z_0$ \footnote{\label{ftnt:fn5} For the  dipole's radiation {\it inside a slab}, then the second term in $G(\tau,0<z<z_0;t,z_0)$,  $\mu^2J_2(\Omega_e\rho)$, which originates from the transmission coefficient $\mathfrak{T}[1/n_e(\nu)]$,  would be absent.}.

Ignoring trivial changes of the arguments, $\tau\to\rho$, $\xi\to\mu$, the result (\ref{eq:E4.9}) for the Green's function coincides with the expression (\ref{eq:E2.8}) for the field of precursor that excites the emission of molecular resonance and is plotted in Fig.2.

\section{ Radiation emitted by excited  molecular resonances: The electric field of radiation.}\label{sec:Sec5}			

The source $\ddot{\cal P}_{mol}$ and the field ${\cal E}_{rad}$ of its radiation are grouped into the terms $\ddot{\cal P}_a+\ddot{\cal P}_b +\ddot{\cal P}_c$ and ${\cal E}_a+{\cal E}_b +{\cal E}_c$, respectively. The source $\ddot{\cal P}$ of a single layer of dipoles located at depth $z_0$ is given by  Eqs. (\ref{eq:E3.7}). The Green's function (\ref{eq:E4.8}b) is used.

\subsection{Structureless (singular)  term ${\cal E}_a $ and regular  term ${\cal E}_b $ } \label{subSec:sbsec5A}

The terms  ${\cal E}_a (\tau,z<0)$ and  ${\cal E}_b(\tau,z<0)$ originate from the $\ddot{\cal{P}}_a$ and $\ddot{\cal{P}}_b$ parts of polarization, respectively. It is shown below that they do not contribute to the total radiation. The singular, $\ddot{\cal{P}}_{a}$, part  of the source  is given by Eq.(\ref{eq:E3.7}a). Using Eq.(\ref{eq:E4.3}) for the backward emitted  part ${\cal E}_a (\tau,z<0|z_0)$, of the electric field yields,
\begin{eqnarray}\label{eq:E5.1}
{{\sf d}{\cal E}_a(\tau,z<0)\over {\sf d}(\Omega_e\tz_0)} =\int_{mT+\tz_0}^{\tau-|\tz|-\tz_0}G(\tau,z<0;t,z_0) ~ 4\pi\ddot{\cal{P}}_a(t|z_0) {d(\Omega_e t) \over\Omega_e^2}
~~~~~~~~~~~~~~~~ ~~~~~~~~~~~~~~~~~~~~~~\nonumber\\
= { \Omega_q^2 \over\Omega_e^2}   \int_{t^*_{min}(m)}^{t^*_{max}}  \sum_{m=0}^{m_p(t)}(-1)^m \epsilon_m G(\tau,z<0;t_*,z_0)  E'_t(t_*-mT) d(\Omega_e t_*),
\end{eqnarray}
where $\Omega_q^2=4\pi q^2 N_q/M$, $t^*_{min}=mT$ and $t^*_{max}=\tau-|\tz|-2\tz_0$ are the time it takes the incident front to reach the dipole and the time it takes the front of the dipole radiation to reach the point $z$ of observation outside the medium at time $\tau$, respectively.  In the same way, by virtue of (\ref{eq:E3.7}b),
\begin{eqnarray}\label{eq:E5.2}
{{\sf d}{\cal E}_b(\tau,z<0)\over{\sf d}(\Omega_e\tz_0)} =\int_{mT+\tz_0}^{\tau-|\tz|-\tz_0} G(\tau,z<0;t,z_0) \cdot 4\pi\ddot{\cal P}_b(t|z_0){d(\Omega_e t)\over\Omega_e^2}
= - { \Omega_q^2 \over\Omega_e^2}    \sum_{m=0}^{m_p(t)} (-1)^m \epsilon_m\int_{t^*_{min}(m)}^{t^*_{max}} d(\Omega_e t_*)    G(\tau,z<0;t_*,z_0)\nonumber\\
\times \omega_0 \int_{mT}^{t_*}dt' e^{-\Gamma_0(t_*-t')}\bigg[(1-{\Gamma_0^2\over\omega_0^2})\sin[\omega_0(t_*-t')]
+2{\Gamma_0\over\omega_0}\cos[\omega_0(t_*-t')]\bigg]E^\prime_t(t'-mT). ~~~~~~
\end{eqnarray}
Here, according to (\ref{eq:E2.8}) and  (\ref{eq:E4.8}b),
\begin{eqnarray}\label{eq:E5.3}
E'_t(t_*-mT)={-{\cal E}_0\over 2\pi i} \oint_{C^-_\omega}{d\uom\over\uom}{\mathfrak T}[n_e(\omega)] \cdot e^{-i\omega(t_*-mT)} e^{-i[\omega-\omega n_e(\omega)]\tz_0}
= {\cal E}_0[J_0(\Omega_e\tau_m)+\gamma_m^2 J_2(\Omega_e\tau_m)],~~~~~~\\ \label{eq:E5.4}
G(\tau,z<0;t_*,z_0)={-i\over 4\pi}\oint_{C_\nu^-}{d\nu\over \nu} \mathfrak{T}[n_e(\nu)] \cdot  e^{-i\nu(\tau-t_*-|\tz|-2\tz_0 )} e^{-i[\nu-\nu n_e(\nu)]\tz_0} =   [J_0(\Omega_e\rho)+\mu^2J_2(\Omega_e\rho)]/2~,~~~~~
\end{eqnarray}
where  $\tau_m^2=(t_*-mT)(t_*-mT+2\tz_0)$, $\gamma_m^2=(t_*-mT)/(t_*-mT+2\tz_0)$ and
$\rho^2=(\tau-t_*-|\tz|-2\tz_0)(\tau-t_*-|\tz|)$, $\mu^2=((\tau-t_*-|\tz|-2\tz_0)/(\tau-t_*-|\tz|)$.  Noteworthy,  the  incident field (\ref{eq:E5.3})  is, in fact, the Green's function, which transforms an incident field that hits the interface, into the field (\ref{eq:E2.8}) of precursor. The Green's function (\ref{eq:E5.4}) differs from the  latter  only by replacement $t\to\tau-t-|\tz|$; it transforms the field of the dipole radiation into the wave outside  medium. Notably, there is no dependence on the parameters $\omega_0$ and $\Gamma_0$ of the molecular dipoles.

To compute the integrals (\ref{eq:E5.1}) and (\ref{eq:E5.2}) we will use the integral representations (\ref{eq:E5.3}) for $G(\tau,z<0;t_*,z_0)$ and $E'_t(t_*-mT)$. Splitting sine and cosine in Eq.(\ref{eq:E5.2})  into two exponents, and integrating $dt'$, we find that
\begin{eqnarray}\label{eq:E5.5}
e^{i\omega mT-\Gamma_0t_*}\int_{mT}^{t_*}
e^{-i(\omega+i\Gamma_0)t'}  e^{\pm i\omega_0(t_*-t')}dt'={i\over \omega+i\Gamma_0\pm\omega_0}
[e^{-i\omega(t_*-mT)} -e^{i(\pm\omega_0+i\Gamma_0)(t_*-mT)}].
\end{eqnarray}
Treated as functions of complex variable $\omega$, these functions are regular and have no pole at the points $\omega=\mp \omega_0 -i\Gamma_0$. The second term in brackets, which comes from the lower limit of the integration, does not depend on $\omega$ and the entire exponent in the integral (\ref{eq:E5.3}) over contour $C_\omega^-$ is reduced to $e^{-i[\omega-\omega n_e(\omega)]\tz_0}$. Its contribution to the contour integral equals to zero. Indeed, after conformal transformation (\ref{eq:E2.5}), the contour $C_\omega^-$ becomes a circle around the origin in $\zeta$-plane, while the exponent  becomes a regular function $e^{-i[\omega-\omega n_e(\omega)]\tz_0}\to e^{-i\Omega_e\tz_0\zeta}$.
Conversely, the exponent stemming from the first term brings in the factor $e^{-i\omega(t_*-mT)}\to e^{-i\Omega_e(t_*-mT)(1/\zeta+\zeta)/2}$, which has an essential singularity at $\zeta=0$.
Assembling  the exponents back into sine and cosine and omitting the $\omega$-independent exponent in brackets yields,
\begin{eqnarray}\label{eq:E5.6}
e^{i\omega mT-\Gamma_0t_*}\int_{mT}^{t_*}
e^{-i(\omega+i\Gamma_0)t'} \stirling {\cos[\omega_0(t_*-t')]}{\sin[\omega_0(t_*-t')]}dt'=
r(\omega)e^{-i\omega(t_*-mT)}\stirling{i\omega-\Gamma_0}{-\omega_0},
\end{eqnarray}
where $r(\omega)=[(\omega+i\Gamma_0)^2-\omega_0^2]^{-1}$ is the resonance factor. As shown above, residues at its poles in the $\omega$ plane are zero.

Using spectral representation (\ref{eq:E5.4}) for the Green's function, we can cast Eqs.(\ref{eq:E5.1}) and (\ref{eq:E5.2})  into two similar double spectral integrals,
\begin{eqnarray}\label{eq:E5.7}
{{\sf d}{\cal E}_a(\tau,z<0)\over{\sf d}(\Omega_e\tz_0)}={\Omega^2_q {\cal E}_0 \over2\pi i\Omega_e}
\sum_{m=0}^{m_p}\epsilon_m (-1)^m
\bigg({-i\over 4\pi }\bigg)\oint_{C_\nu^-}{d\nu\over \nu} \mathfrak{T}[n_e(\nu)]e^{-i\nu(\tau-|\tz|)}
e^{+i[\nu+\nu n_e(\nu)]\tz_0}\nonumber\\
\times \oint_{C_\omega^-}{d\omega\over\omega} \mathfrak{T}[n_e(\omega)] e^{i\omega m T}  e^{-i[\omega-\omega n_e(\omega)]\tz_0}
\cdot \int_{t^*_{min}(m)}^{t^*_{max}}  e^{i (\nu-\omega) t_*} dt_*. ~~~~~
\end{eqnarray}
\begin{eqnarray}\label{eq:E5.8}
{{\sf d}{\cal E}_b(\tau,z<0)\over{\sf d}(\Omega_e\tz_0)}=-{\Omega^2_q {\cal E}_0 \over2\pi i\Omega_e}
\sum_{m=0}^{m_p}\epsilon_m (-1)^m
\bigg({-i\over 4\pi }\bigg)\oint_{C_\nu^-}{d\nu\over \nu} \mathfrak{T}[n_e(\nu)]e^{-i\nu(\tau-|\tz|)}
e^{+i[\nu+\nu n_e(\nu)]\tz_0}\nonumber\\
\times \oint_{C_\omega^-}{d\omega\over\omega} \mathfrak{T}[n_e(\omega)]r(\omega)[2i\Gamma_0\omega-\omega_m^2] e^{i\omega m T}  e^{-i[\omega-\omega n_e(\omega)]\tz_0}
\cdot \int_{t^*_{min}(m)}^{t^*_{max}}  e^{i (\nu-\omega) t_*} dt_*. ~~~~~
\end{eqnarray}
The last integral in the above equations is readily found to be
\begin{eqnarray}\label{eq:E5.9}
\int_{t^*_{min}(m)}^{t^*_{max}}  e^{i (\nu-\omega) t_*} dt_= {-i\over \nu-\omega} \bigg[e^{i(\nu-\omega)(\tau-|\tz|-2\tz_0)}-	e^{i(\nu-\omega)mT}\bigg],
\end{eqnarray}
Therefore, we can rewrite  equations  (\ref{eq:E5.7}) and   (\ref{eq:E5.8})  as
\begin{eqnarray}\label{eq:E5.10}
{{\sf d}{\cal E}_a(\tau,z<0)\over{\sf d}(\Omega_e\tz_0)}={\Omega^2_q {\cal E}_0 \over 2\pi i\Omega_e}
\bigg({-i\over 4\pi }\bigg)\oint_{C_\nu^-}{d\nu\over \nu} \mathfrak{T}[n_e(\nu)]e^{-i[\nu-\nu n_e(\nu)]\tz_0} \oint_{C_\omega^-}{d\omega\over\omega} \mathfrak{T}[n_e(\omega)]e^{-i[\omega-\omega n_e(\omega)]\tz_0}
~~~~~~~~~~~~~~~~~~~~~~~\nonumber\\
\times \sum_{m=0}^{m_p}\epsilon_m (-1)^m {e^{-i\omega (\tau-|\tz| -2\tz_0-m T)}
	- e^{-i\nu(\tau-|\tz| -2\tz_0-m T)}\over \nu-\omega},~~~~~~~~~~~~~~~~~~~~~~~~~~~~~~
\end{eqnarray}
\begin{eqnarray}\label{eq:E5.11}
{{\sf d}{\cal E}_b(\tau,z<0)\over{\sf d}(\Omega_e\tz_0)}=-{\Omega^2_q {\cal E}_0 \over 2\pi i\Omega_e} \bigg({-i\over 4\pi }\bigg)\oint_{C_\nu^-}{d\nu\over \nu}
\mathfrak{T}[n_e(\nu)]e^{-i[\nu-\nu n_e(\nu)]\tz_0} \oint_{C_\omega^-}{d\omega\over\omega} \mathfrak{T}[n_e(\omega)] r(\omega)[2i\Gamma_0\omega-\omega_m^2]~e^{-i[\omega-\omega n_e(\omega)]\tz_0}\nonumber\\
\times \sum_{m=0}^{m_p}\epsilon_m (-1)^m    {e^{-i\omega (\tau-|\tz| -2\tz_0-m T)}
	- e^{-i\nu(\tau-|\tz| -2\tz_0-m T)}\over \nu-\omega}~~~~~~~~~~~~~~~~~~~~~~~~~~~~~~
\end{eqnarray}
The integrands, as functions of two complex variables $\omega$ and $\nu$, have a removable singularity at $\omega=\nu$ in either of the two complex planes and the residue in these poles equal zero.
Further analytic calculations, which are described in Appendix C, show that
\begin{eqnarray}\label{eq:E5.12}
{{\sf d}{\cal E}_a(\tau,z<0)\over{\sf d}(\Omega_e\tz_0)}={{\sf d}{\cal E}_b(\tau,z<0)\over{\sf d} (\Omega_e\tz_0)}=0,
\end{eqnarray}
This result could have been anticipated from the viewpoint of the rigorous theory of dispersion \cite{Born,Rosenfeld}. Indeed, the terms ${\cal P}_a$ and ${\cal P}_b$ represent a linear polarization response of the molecular resonances  to the external fields  of the {\em entire train} (\ref{eq:E3.3}) of primary precursors with zero initial conditions, $X(\tz_0|z_0)=\dot{X}(\tz_0|z_0)=0$, which is equivalent to Eqs. (\ref{eq:A.3}).  The $m$-dependent limits of integration in Eqs. (\ref{eq:E5.1}) and (\ref{eq:E5.2}) come from the theta-functions $\theta(t_\ast-mT)$ in Eq.(\ref{eq:E3.3}). The absence of backward radiation from the
components  ${\cal P}_a$ and ${\cal P}_b$ indicates that  they  satisfy the homogeneous wave equation for the average polarization ${\cal P}$  with a refraction index $n(\omega)$, which is the  {\em main postulate} of molecular optics (see  \S 2.4 of the textbook by M. Born and E. Wolf \cite{Born} and, especially, the  Ch.VI of  lectures by L. Rosenfeld \cite{Rosenfeld}) .

\subsection{Oscillatory  term ${\cal E}_c$}\label{subSec:sbsec5B}

	With the $\ddot{\cal{P}}_c$ defined by Eq.(\ref{eq:E3.7}c), and the Green's function (\ref{eq:E4.8}b) the component ${\cal E}_c$ of the radiation field reads as
	\begin{eqnarray}\label{eq:E5.13}
	{{\sf d}{\cal E}_c(\tau,z<0|z_0)\over (\Omega_e {\sf d}\tz_0)}=\int G(\tau,z;t,z_0) \cdot \Omega_e^{-2} 4\pi\ddot{\cal P}_c(t|z_0) d(\Omega_e t)
	= 4\pi{ \uom_0^2}\int_{t_{min}(m_p)}^{t_{max}}d(\Omega_e t_*)  e^{-\Gamma_0(t_*-m_pT)}
	~~~~~\nonumber\\
	\times {-i\over 4\pi}\oint_{C_\nu^-}{d\nu\over \nu} \mathfrak{T}[n_e(\nu)] \cdot  e^{-i\nu(\tau-t_*-|\tz|)}e^{+i[\nu +\nu n_e(\nu)]\tz_0} \bigg[C_1\cos[\omega_0 (t_*-m_pT)]
	+ C_2 \sin[\omega_0 (t_*-m_pT)]   \bigg] ,
	\end{eqnarray}
	where coefficients $C_1(m_pT)$ and $C_2(m_pT)$ are defined in Eqs. (\ref{eq:E3.8}) and, according to (\ref{eq:A.8}), all the ${\cal P}(m_p T)$ carry the same dimensionless factor $\Omega_q^2/\Omega_e^2$.
	The final integration $dt_*$ is carried out after the ladder of amplitudes of free oscillations from $m=0$ to $m=m_p$ is built according to the recursive relations (\ref{eq:A.7}).
	Taking the Green's  function in the form (\ref{eq:E4.8}b), we integrate over $t_*$ between $t^*_{min}=m_pT$ and $t^*_{max}=\tau-|\tz|-2\tz_0$,
	\begin{eqnarray}\label{eq:E5.14}
	e^{+\Gamma_0m_pT}\int_{t^*_{min}}^{t^*_{max}} e^{i(\nu+i\Gamma_0 )t_*}[(C_1-iC_2) e^{i\omega_0 (t_*-m_pT)} +(C_1+iC_2)e^{-i\omega_0 (t_*-m_pT)}]dt_*
	\end{eqnarray}
    Assembling the result (\ref{eq:C.5}) of integration into Eq.(\ref{eq:E5.13}) we arrive at
	\begin{eqnarray}\label{eq:E5.15}
	{{\sf d}{\cal E}_c(\tau,z<0|z_0)\over \Omega_e{\sf d}\tz_0}\!=\!{4\pi\uom_0^2\Omega_e^2\over 4\pi i}
	\!\!\oint_{C_\nu^-}\!\!{d\nu\over \nu} \mathfrak{T}[n_e(\nu)] r(\nu) 	e^{-i[\nu-\nu n_e(\nu)]\tz_0}
	\bigg[  [\ldots]\!-\! e^{-i \nu \tau_*}[(i\unu -\uGa_0)C_1(m_pT)\!-\! \uom_0C_2(m_pT)]\bigg]\!,~~~~~
	\end{eqnarray}
	where $r(\nu)=[(\nu+i\Gamma_0)^2-\omega_0^2]^{-1}$, $\tau_*=t^*_{max}-t^*_{min}=\tau-|\tz|-2\tz_0-m_pT$ and expression $[\ldots]$ in brackets (originating from the upper limit of integration in Eq.(\ref{eq:E5.14})) is a linear combination of the products like $C_{1,2}e^{-\Gamma_0\tau_*}e^{\pm i\omega_0\tau_*}$. This term does not contain powers of $\nu$ higher than one.	Here, the only $\nu$-dependent exponent is $e^{-i[\nu-\nu n_e(\nu)]\tz_0}$. Hence,
	$$ \oint_{C_\nu^-}{d\nu\over \nu} \mathfrak{T}[n_e(\nu)] r(\nu)	e^{-i[\nu-\nu n_e(\nu)]\tz_0}~\to~\oint^{(0+)}   {d\zeta \over \zeta}  (1-\zeta^2)~r(\zeta)e^{-i\Omega_e\tz_0\zeta}~=0,$$
    and this integral will turn to zero after  integration over the contour (cf. Appendix{ \ref{app:appC}}). However, the  terms associated with the lower limit, where the product of  exponents, $ e^{-i[\nu-\nu n_e(\nu)]\tz_0} e^{-i\nu(\tau-|\tz|-2\tz_0-m_pT)}$, has an essential singularity in the $\zeta$-plane, must be retained.
	Therefore,
	\begin{eqnarray}\label{eq:E5.16}
	{{\sf d}{\cal E}_c(\tau,z<0|z_0)\over {\sf d}(\Omega_e\tz_0)}\!
	={4\pi\uom_0^2\Omega_e^2\over 4\pi i}
	\oint_{C_\nu^-}\!{d\nu\over \nu} \mathfrak{T}[n_e(\nu)] r(\nu) e^{-i\nu(\tau-|\tz|-2\tz_0-m_pT)}
	e^{-i[\nu-\nu n_e(\nu)]\tz_0]}  [\uom_0 a_1(m_pT) -i\unu a_2(m_pT)],~~~~~
	\end{eqnarray}
	where
	\begin{eqnarray}\label{eq:E5.17}
	a_1(m_pT)= \bigg(1+{\Gamma_0^2\over\omega_0^2}\bigg){\dot{\cal P}(m_p T)\over\omega_0},~~
	a_2(m_pT)= \bigg(1+ {\Gamma_0^2\over\omega_0^2}\bigg){\cal P}(m_p T)
	+ 2{\Gamma_0\over \omega_0}{\dot{\cal P}(m_p T)\over\omega_0}.
	\end{eqnarray}
	In terms of variable $\zeta$ (cf. Eq.(\ref{eq:E2.5})) and with the resonance factor $r(\zeta)$ given by Eq.(\ref{eq:B.2}), Eq.(\ref{eq:E5.16}) reads as
	\begin{eqnarray}\label{eq:E5.18}
	{ {\sf d}{\cal E}_c(\tau,z<0|z_0)\over {\sf d}(\Omega_e\tz_0)}= {4\pi\uom_0\over 2\pi i} \oint^{(0+)}{d\zeta\over \zeta} \exp{\bigg\{-i{\Omega_e\Lambda\over 2}\bigg[ {\Xi\over \zeta}+{\zeta\over\Xi}\bigg]\bigg\}}~~~~~~~~~~~~~~~~~~~~~~~~~~~~~~~~~~~~~~~~~~~\\
	\times\sum_{l=1}^{\infty}\bigg[{\rm Re} \bigg({\sin2l\vartheta\over\sin\vartheta}\bigg)\zeta^{2l}
	+ i{\rm Im} \bigg({\sin(2l+1)\vartheta\over\sin\vartheta}\bigg)\zeta^{2l+1}\bigg]
	\bigg\{\uom_0a_1(m_pT)(1-\zeta^2) -a_2(m_pT){i\over 2}(\zeta^{-1}-\zeta^3 )\bigg\},\nonumber
	\end{eqnarray}
	where $\Lambda^2=(\tau-|\tz|-m_pT-\tz_0)^2-\tz_0^2$ and  $\Xi^2=(\tau-|\tz|-m_pT-2\tz_0)/(\tau-|\tz|-m_pT)$.
	Using the integral representation (\ref{eq:E2.7}) of the Bessel coefficients,
	\begin{eqnarray}\label{eq:E5.19}
	{ {\sf d}{\cal E}_c(\tau,z<0|z_0)\over {\sf d}(\Omega_e\tz_0)}= {{\cal E}_0\Omega_q^2\over\Omega_e^2}\uom_0 \bigg\{\uom_0a_1(m_pT)[s_1(\Lambda,\Xi)+s_3(\Lambda,\Xi)] -{a_2(m_pT)\over 2} [s_2(\Lambda,\Xi)+s_4(\Lambda,\Xi)]\bigg\},
	\end{eqnarray}
	where  the factor ${\cal E}_0\Omega_q^2/\Omega_e^2$, which, starting from  (\ref{eq:A.3}), is present in every function $4\pi a_j(m_pT)$, is factored out. The functions $s_j(\Lambda,\Xi)$, which are defined by Eqs.(\ref{eq:D.1}) are the sums of  products, $\sum_{l}\Xi^{2l} J_{2l}(\Omega_e\Lambda)$. Their behavior critically depends on the distance $z_0$ to the interface and cannot be comprehended without foregoing analysis of primary precursors' propagation presented in Sec.\ref{sec:Sec2}. This is addressed in detail in Sec.\ref{subsec:Sec6A} and Appendix \ref{app:appD}.

\section{ Spectroscopy with precursors, results of calculations and  discussion}\label{sec:Sec6}	

\subsection{Results of calculations}\label{subsec:Sec6A}

The main results of this study are represented by Eq.~(\ref{eq:E5.19}). This equation reflects an explicit dependence of the field radiated in the backward direction
on the time interval, $\Delta t =\tau-|\tz|-m_pT$,  it takes the $m_p$-th pulse to travel from the interface to a detector located at some distance $z$  from vacuum-medium  interface, where $T$  is the duration of an individual pulse. The field (\ref{eq:E5.19})  substantially depends on the depth $z_0$ of a dipole's location in a medium. The expression in curly brackets is the sum of two terms each of which is a product of an $m_p$-dependent amplitude and the time-dependent signal.
\begin{figure}[h]
	\includegraphics[width=0.9\textwidth]{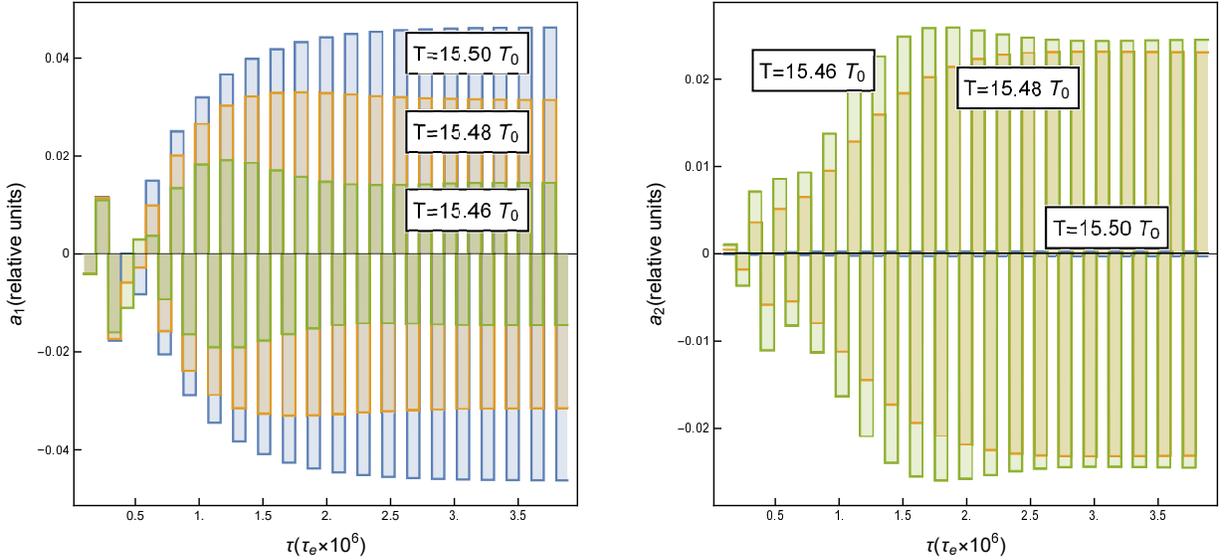}
	\caption{Time dependence of ladder coefficients $a_1$ and $a_2$. Amplitudes of the ladder parts of the signal level out with time at any value of $T$. The maxima of combined signals determine the resonant values for $T$.}
	\label{Fig:fig5}
\end{figure}

\begin{figure}[t]
	\includegraphics[width=0.9\textwidth]{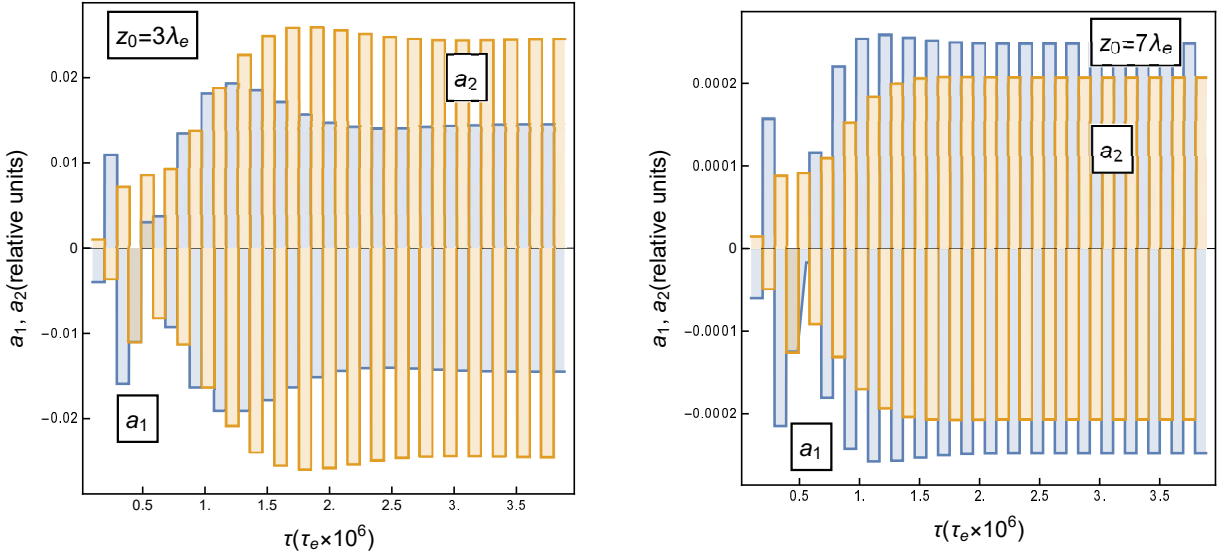}
	\caption{Comparison of ladder amplitudes for two values of $z_0$, $3\lambda_e$ and $7\lambda_e$. The amplitudes of the ladder decrease by the factor of 100 as the depth of the radiating layer increases from $3\lambda_e$ to $7\lambda_e$. }
	\label{Fig:fig6}
\end{figure}

The time dependence of $a_1(m_pT)$ and  $a_2(m_pT)$  shown in Fig.~\ref{Fig:fig5}, initially increase and eventually levels off forming a shape that resembles a ladder. When $\Gamma_0\ll\omega_0$ then, according to Eqs.(\ref{eq:E5.17}), parameters $a_1(m_pT)\propto\dot{\cal P}(m_p T)$ and $a_2(m_pT) \propto{\cal P}(m_p T)$ approximately coincide with deviations of velocity and displacement  from their zero  values at equilibrium the moment when the $m_p$-th pulse hits the dipoles. The precise timing of a hit is critical for attaining an amplification of the amplitude of the dipole oscillation.  The maximum velocity after a hit  is attained when the duration of a pulse is $T=(n+1/2)T_0$ , which is clearly seen in Fig.~\ref{Fig:fig5}  (where $n=15$). This hit coincides with the dipole passing through its equilibrium position and ${\cal P}(m_p T)\approx 0$. The added $1/2$ accounts for the change of polarity of  successive pulses. Even a small deviation from the resonant value of $T$ results in a visible decrease of the amplitude  of velocity, $\propto a_1$, and increase  of the coordinate, $\propto a_2$, so that $a_1$ and $a_2$ become comparable. Saturation of both amplitudes with growth of $m_p$ is observed for all $T$.  All calculations shown in Fig.~\ref{Fig:fig5} are done with $\Omega_e=10^{15}$~rad/sec, $\omega_0=10^{12}$~rad/sec, $\Gamma_0=4\times10^9$ rad/sec, and $z_0=3\lambda_e$.
For larger $z_0$, the behavior of the ladder amplitudes remains qualitatively the same except that their values decrease by orders of magnitude. This is illustrated in Fig.\ref{Fig:fig6} where $a_1$ and $a_2$ are plotted for $z_0=3 \lambda_e$ and $z_0=7\lambda_e$ with such a non-resonant value of $T$, that  $a_1$ and $a_2$  are comparable,

\begin{figure}[h]
	\includegraphics[width=0.5\textwidth]{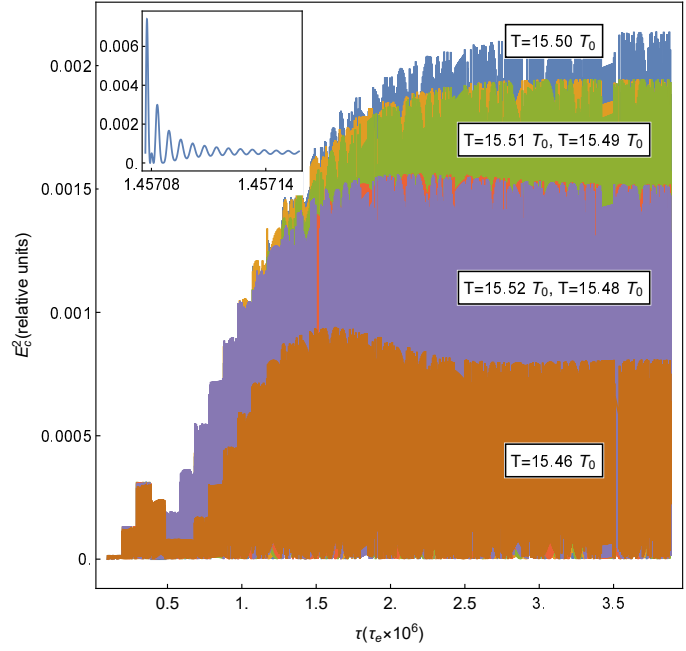}
	\caption{Square of radiated field from a single layer at $z_0=3\lambda_e$ for different values of $T$ in the vicinity of resonant $T=15.5T_0$. The saturation of amplitude of intensity of the backward radiation is maximized at resonant values of $T$, the resonances are rather sharp. Different colors (or shades of gray) show the approach to resonance from $T=15.46T_0$ to  $T=15.50T_0$; and then a symmetric drop to  $T=15.52T_0.$ }
	\label{Fig:fig7}
\end{figure}

In the final answer (\ref{eq:E5.19}) for the electric field ${\cal E}_c(\tau,z<0|z_0)$ of radiation, the ladder amplitudes $a_1$ and $a_2$ are multiplied by the oscillating source functions $s_1+s_3$ and $s_2+s_4$, respectively.
The square of this field (proportional to the radiated power), calculated for several values of $T$, is shown in Fig.~\ref{Fig:fig7} for a small $z_0=3\lambda_e$.
Here, $T=15.50 T_0$ is a resonant value of $T$.  The general condition for the resonance, $T_{res}=(n+1/2)T_0$, provides an opportunity to measure $\omega_0$; the difference between adjacent resonances in $T$ is equal to $T_0=2\pi/\omega_0$.
The  beginning of each pulse is accompanied by the precursor, shown in the inset of Fig.\ref{Fig:fig7}.
\begin{figure*}[t]
	\includegraphics[width=0.9\textwidth]{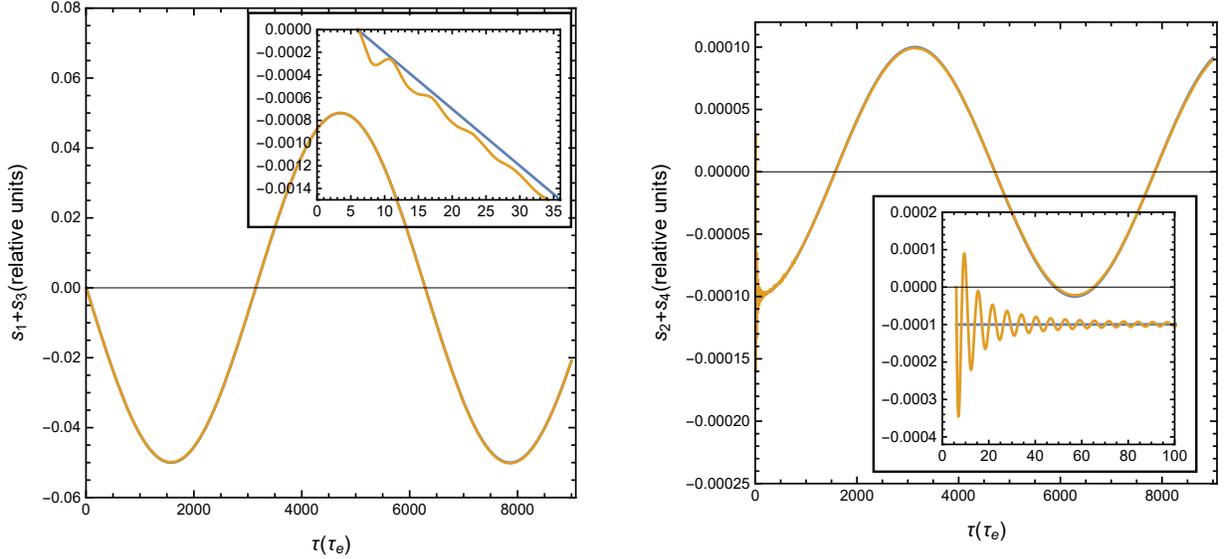}
	\caption{Time dependence of signals $s_1+s_3$ and $s_2+s_4$ for shallow dipoles ($z_0=3 \lambda_e$). Left panel: on a large scale the signal (blue) coincides with a harmonic component (orange); on a small scale shown in the inset, the difference due to the precursor is visible at an early time. 	Right panel: same features as on the left panel, except that in $s_2+s_4$ the precursor is more pronounced.}
	\label{Fig:fig8}
\end{figure*}
The deeper the molecular dipole is located, the smaller is the amplitude of the associated harmonic oscillations and the more pronounced are the secondary precursors of its backward radiation, which is triggered by the sharp fronts of the primary precursors.
\begin{figure*}[t]
	\includegraphics[width=0.9\textwidth]{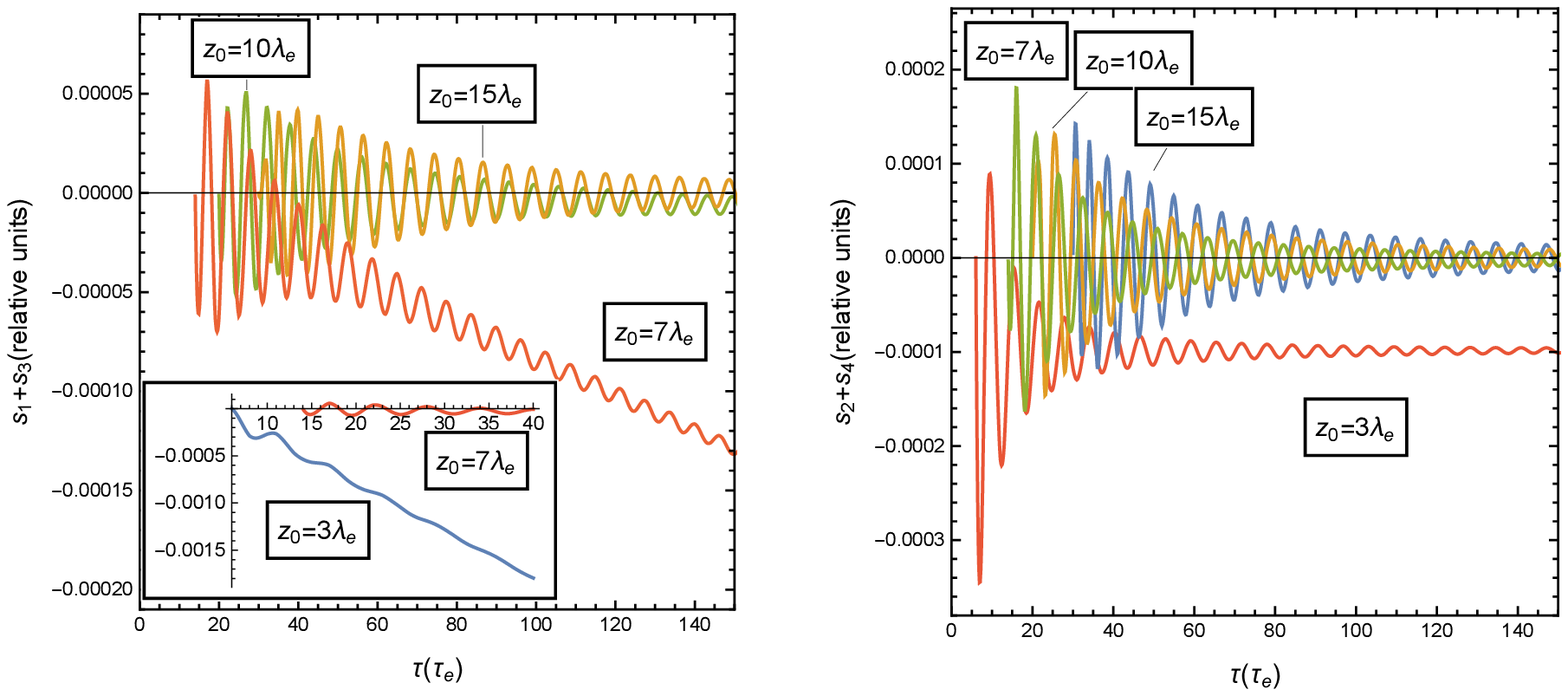}
	\caption{Comparison of time dependencies of signals $s_1+s_3$ (left panel) and $s_2+s_4$(right panel) for different values of $z_0$. Each signal is a sum of a harmonic signal (negative sine on the left and negative cosine on the right) and an oscillating precursor. As the depth increases, amplitudes of harmonic parts sharply decrease and the precursor parts (starting from the same amplitude) attenuate less and less.}
	\label{Fig:fig9}
\end{figure*}
This can be seen in Figs.~\ref{Fig:fig8} and~\ref{Fig:fig9}, were we plot, with the same parameters, signals $s_1+s_3$ and $s_2+s_4$ for shallow, $z_0=3\lambda_e$, and deep, $z_0\geq 7\lambda_e$, dipoles, respectively. Every $s_j$ appears to be a sum of a harmonic signal originating from an oscillating dipole and a {\em secondary precursor} formed by the electronic polarization near the leading front.
The pattern can be qualitatively understood from the dependence of primary precursors on the distance its leading front has penetrated into the  medium, as given by Fig.\ref{Fig:fig2}.  For shallow dipoles, the impulse obtained from the relatively smooth electric field of primary precursors is large and the dipoles immediately begin a harmonic motion, as in Fig.\ref{Fig:fig8}. For deeper dipoles, the first peak of the electric field of primary precursors is too sharp to excite harmonic oscillations of large amplitude. Instead, a secondary precursor is produced, as in Fig.\ref{Fig:fig9}. This behavior is confirmed by the asymptotic formulae  (\ref{eq:D.15})-(\ref{eq:D.18}),  which bear the pattern $J_0(\Omega_e\tau)-\cos(\omega_0\tau)$, where a molecular harmonic and a precursor are clearly visible.

\subsection{Discussion}\label{subsec:Sec6B}

Although we intend to measure the same characteristics of matter that are traditionally studied by means of spectroscopy, our approach is quite different. We propose to probe properties of matter by a train of square pulses with sharp wavefronts. This approach does not rely on any kind of spectral device and is motivated by an inherently large difference in scales of the physical processes that result in an actually observed  signal and prompt its interpretation. These scales are associated with the Langmuir frequency $\Omega_e\sim 10^{15} - 10^{16} $ rad/sec of the electronic component of polarization, the proper frequency $\omega_0\sim 10^{12}$ rad/sec  and the width $\Gamma_0$ of a molecular resonance, and, finally, frequency $\nu_0\sim 10^{8}-10^{10}$/sec of the pulses' repetition in the incident train that is used to probe molecular resonances. Each of these processes allows for an exact analytical treatment.

The first process is the formation of  {\em primary precursors} at the vacuum-medium interface. It depends only on the highest frequency $\Omega_e$ (\ref{eq:E2.1}), which is translated into the finest time interval $\tau_e=2\pi/\Omega_e$ and the shortest distance $\lambda_e=c\tau_e$. Its sole parameter is the density of {\em all electrons} in a medium. The electric field of primary precursors can be found analytically as light electrons begin to radiate and develop collective behavior forming an index of refraction nearly immediately. Therefore, this stage is totally under the jurisdiction of the rigorous theory of dispersion \cite{Rosenfeld,Born}. Our calculations show, that in the vicinity of the interface ($z\lesssim 5\lambda_e$), the electric field near the leading front smoothly oscillates. But the deeper the front penetrates inside a medium, the sharper the first oscillations become. Regardless of how deeply the pulse penetrates a medium, its amplitude at the leading front stays the same as the amplitude of the incident signal.

The second process is the excitation of oscillations in heavy molecular dipoles.  The field acting on heavy elastic dipoles is not an incident monochromatic wave or a train of square pulses, but rather a field of  precursors formed by the electronic component of polarization.  It  takes many periods $T_0=2\pi/\omega_0$ of proper oscillations to develop a collective behavior of molecular dipoles that would have contributed to the  refraction index. In our case this limit is {\em not} reached. Instead, we address the problem of driving the proper oscillations by a train of primary precursors directly. We solve the equation of motion (\ref{eq:E3.1}) for heavy elastic dipole, located at a distance $z_0$ from the interface, in the presence of the electric field (\ref{eq:E3.3}) of the primary precursors. Within the model of a classical oscillator, this problem allows for an analytic solution. We predict that, starting from the first pulses, the amplitudes of dipole's oscillation would form an ascending ladder  that  eventually reaches some saturation level, unless the damping $\Gamma_0=0$.  The former must be maximum when the duration $T=1/\nu$ of an individual incident pulse is $T_n=(n+1/2)T_0=2\pi(n+1/2)/\omega_0$, which indicates the existence of a resonance. The time interval, $T_0$, between two neighboring resonances determines the frequency  of proper oscillations, $\omega_0$.

The third process is the emission of secondary radiation by the oscillating molecular dipoles. We have shown that, in the final answer for the measured signal, the ladder amplitudes are multiplied by the time-dependent {\em source functions}, which explicitly depend on time and the distance  of a radiating dipole from the interface, $z_0$. Numerical analysis confirms these functions to be the sums of two distinctive parts. The first part is a harmonic function oscillating with the proper frequency $\omega_0$ of a molecular resonance. It dominates for the dipoles located close to the interface, $z_0\lesssim 10\lambda_e$, and is due to the oscillations excited by the leading and relatively smooth parts of primary precursors. The second part  of the radiated field oscillates with the Langmuir frequency $\Omega_e$  and represents the train of {\em secondary precursors}, which propagate not only forward, but also in the {\em backward direction} with respect to the  incident train. Secondary precursors come predominantly from deeply located dipoles, where the near-front oscillations of primary precursors are very fast. This could have been anticipated from a qualitative inspection of Fig.\ref{Fig:fig2},

All numerical calculations were performed with the Wolfram's {\em Mathematica } software on a standard PC. Because the Lommel's functions are composed from highly oscillating Bessel functions, their numerical implementation meets difficulties. To work around them, we developed a special procedure, which is explained in Appendix~\ref{app:appD}. More accurate calculations may require faster computers.

Finally, we would like to discuss the prospect of building a device implementing the proposed here idea of precursor-based spectroscopy.  Generators that are able to deliver a train of square pulses with the repetition frequency of 1-5 GHz are widely available. The harmonic part of the backward radiation with frequency about a few terahertz can be rectified, e.g., using a Schottky diode as a power detector, and measured with a reasonable precision. The most intriguing possibility may exist due to the secondary precursors in the backward radiation. These can be used to synchronize the incident signal with the returned signal, possibly, even forming a standing wave between the generator and a sample.

	\appendix
	
	\section{Calculation of the ladder of excitations}
		\label{app:appA}
	\renewcommand{\theequation}{A.\arabic{equation}}
	\setcounter{equation}{0}
		Let $m_p =m_p(t_\ast)$ be the number of wavefronts that have crossed $z_0$ by the time $t$, $m_pT \leqslant t<(m_p+1)T$ so that $m_p=0$ corresponds to the leading front that enters a yet not polarized medium. When the dipole is hit by the next pulse, the upper limit $m_p(t_*)$ increases by one. In order to compute $X(t|z_0)$, we substitute (\ref{eq:E3.3}) into Eq.(\ref{eq:E3.5}), which yields,
		\begin{eqnarray}\label{eq:A.1}
		X(t|z_0)={{\cal E}_0 q \over M}  \cdot \int_{0}^{t_\ast} K(t_\ast-t^\prime_\ast)
		\bigg[  \sum_{m=0}^{m_p(t_\ast)}\epsilon_m(-1)^m   \theta(t^\prime_\ast-mT)E^\prime_t(t^\prime_\ast-mT)\bigg] d t^\prime_\ast~,
		\end{eqnarray}
		where kernel $K(t-t^\prime)$, the fundamental solution of differential equation (\ref{eq:E3.1}), and its first two time derivatives are as follows,
		\begin{eqnarray}\label{eq:A.2}
		K(x)&=& \theta(x)e^{-\Gamma_0 x}{\sin\omega_0 x\over \omega_0}, ~~~\dot{K}(x)={dK(x)\over dx}= \theta(x)e^{-\Gamma_0 x} \bigg[\cos\omega_0 x-\Gamma_0 {\sin\omega_0 x\over \omega_0}\bigg],~~~~~~~ \nonumber\\
		\ddot{K}(x)&=&{d^2K(x)\over dx^2}=\delta(x)+ \theta(x)e^{-\Gamma_0 x}\bigg[-2\Gamma_0\cos\omega_0 x
		-(\omega_0^2 -\Gamma_0^2) {\sin\omega_0 x\over \omega_0}\bigg],~~~K(0)=0,~~~\dot{K}(0)=1~.
		\end{eqnarray}
		Since we assume that the dipole's charges are initially at rest, then for $0<t_*<T$,
		\begin{eqnarray}\label{eq:A.3}
		X_{(0)}(t_*)={{\cal E}_0q\over M}\int_0^{t_*} K(t_\ast-t^\prime)E^\prime_t(t^\prime)dt',~~
		\dot{X}_{(0)}(t_*)={{\cal E}_0q\over M}\int_0^{t_*} \dot{K}(t_\ast-t^\prime)E^\prime_t(t^\prime)dt',
		~~X_{(0)}(0)=\dot{X}_{(0)}(0)=0.~~~
		\end{eqnarray}
		If the $m_p$-th pulse is passing through a dipole, then for  $m_pT<t_*<(m_p+1)T$,
		\begin{eqnarray}\label{eq:A.4}
		X_{(m_p)}(t_*)={{\cal E}_0q\over M}\int_{m_pT}^{t_*} K(t_*-t^\prime) [\sum_{m=0}^{m_p(t_*)}(-1)^m\epsilon_m\theta(t^\prime_\ast-mT)E^\prime_t(t^\prime-mT)] dt ' ~~~~~~~~~~~~~~~~~~~~~~~~~~~~~~~~~~~~~~~\nonumber\\
		+e^{-\Gamma_0 t_*}[ b_c(m_pT) \cos\omega_0t_*+ b_s(m_pT) \sin\omega_0t_*],~~~~~~~~~~~~~~~~~~~~~~~~~\\
		\dot{X}_{(m_p)}(t_*)={{\cal E}_0q\over M}\int_{m_pT}^{t_*} \dot{K}(t_*-t^\prime) [\sum_{m=0}^{m_p(t_*)}(-1)^m\epsilon_m\theta(t^\prime_\ast-mT) E^\prime_t(t^\prime-mT)] dt'~~~~~~~~~~~ ~~~~~~~~~~~~~~~~~~~~~~~~~~~~\nonumber\\
		+ e^{-\Gamma_0 t_*}\{[\omega_0b_s(m_pT)-\Gamma_0b_c(m_pT)] \cos\omega_0t_*
		-  [\omega_0b_c(m_pT)+\Gamma_0b_s(m_pT)]\sin\omega_0t_*\}~.~~\nonumber
		\end{eqnarray}
		The coefficients $b_c(m_pT)$ and  $b_s(m_pT)$ can be expressed in terms of the dipole's amplitude $X(m_pT)$ and its time derivative $\dot{X}(m_pT)$.
		Coordinate and velocity at time $T$ are continuous, i.e., their values at the end of the first pulse, $m_p=0$, and at the  beginning of the second pulse, $m_p=1$, are equal. According to (\ref{eq:A.3}), these  are
		\begin{eqnarray}
		X_{(0)}(T)=X_{(1)}(T) ={{\cal E}_0q\over M}\int_0^{T} K(T-t^\prime)E^\prime_t(t^\prime)dt', ~~~
		\dot{X}_{(0)}(T)= \dot{X}_{(1)}(T) ={{\cal E}_0q\over M}\int_0^{T} \dot{K}(T-t^\prime)E^\prime_t(t^\prime)dt'~.\nonumber
		\end{eqnarray}
		Similarly, if in  Eq.~(\ref{eq:A.4}) $t_*=m_pT$, at the {\em beginning} of the $m_p$-th interval, the integrals become zero and
		\begin{eqnarray}\label{eq:A.5}
		X_{(m_p)}(m_pT)=e^{-\Gamma_0m_pT}[ b_c(m_pT) \cos\omega_0m_pT+ b_s(m_pT)\sin\omega_0m_pT], ~~~~~~~~~~~~~~~~~~~~~~~~~~~~~~~~~~~~~~~~~~~~~~~\\
		\dot{X}_{(m_p)}(m_pT)=e^{-\Gamma_0 m_pT} \{[\omega_0b_s(m_pT)-\Gamma_0b_c(m_pT)] \cos\omega_0m_pT
		-  [\omega_0b_c(m_pT)+\Gamma_0b_s(m_pT)]\sin\omega_0m_pT\}.\nonumber
		\end{eqnarray}
		Now, we can trade $b_c(mT)$ and $b_s(mT)$ for $X(mT)$ and $\dot{X}(mT)$:
		\begin{eqnarray}\label{eq:A.6}
		\omega_0e^{-\Gamma_0m_pT}b_c(m_pT)=(\omega_0\cos\omega_0 m_pT-\Gamma_0\sin\omega_0m_pT)\cdot X_{(m_p)}(m_pT) -\sin\omega_0m_pT\cdot \dot{X}_{(m_p)}(m_pT)~,  \nonumber\\
		\omega_0e^{-\Gamma_0 m_pT}b_s(m_pT)=(\omega_0\sin\omega_0m_pT+\Gamma_0\cos\omega_0m_pT)\cdot X_{(m_p)}(m_pT)+\cos\omega_0m_pT\cdot \dot{X}_{(m_p)}(m_pT)~.
		\end{eqnarray}
		Then for $m_p(t_\ast)T<t_\ast<(m_p(t_\ast)+1)T$,
		\begin{eqnarray}\label{eq:A.7}
		X_{(m_p)}(t_*)={{\cal E}_0q\over M}\int_{m_p T}^{t_*} K(t_\ast-t^\prime)~\sum_{m=0}^{m_p(t_*)}(-1)^m\epsilon_m \theta(t'-mT) E^\prime_t(t'-mT)~ dt' \nonumber ~~~~~~~~~~~~~~~~~~~~~~~~~~~~~~~~~~~~~~~~~~~~\\
		+ e^{-\Gamma_0(t_*-m_pT)}\bigg\{\bigg[\cos\omega_0 (t_*-m_p T)
		+{\Gamma_0\over \omega_0}\sin\omega_0 (t_*-m_p T) \bigg]X_{(m_p)}(m_pT)
		+\sin\omega_0 (t_*-m_p T) {\dot{X}_{(m_p)}(m_p T)\over \omega_0}\bigg\} ,~~~~\nonumber\\
		{\dot{X}_{(m_p)}(t_*) \over \omega_0}={{\cal E}_0q\over M}\int_{m_pT}^{t_*} {\dot{K}(t_\ast-t')\over \omega_0} [\sum_{m=0}^{m_p(t_*)}(-1)^m\epsilon_m\theta(t'-mT) E^\prime_t(t^\prime-mT)]dt'~+~ e^{-\Gamma_0(t_*-m_pT)}  ~~~~~~~~~~~~~~~~~\\
		\times\bigg\{-\bigg(1+{\Gamma_0^2\over \omega_0^2}\bigg)\sin\omega_0 (t_*-m_pT)X_{(m_p)}(m_pT)
		+\bigg[\cos\omega_0 (t_*-m_pT)   -{\Gamma_0\over\omega_0}\sin\omega_0 (t_*-m_pT)\bigg]
		{\dot{X}_{(m_p)}(m_p T)\over\omega_0}\bigg\}~~~~~~~ \nonumber
		\end{eqnarray}
		For $t_*=m_pT$, the above equations become identities.
		For $t_*=(m_p+1)T$ in Eqs.(\ref{eq:A.7}), we obtain the recursion formula for coefficients ${X}_{(m_p)}(m_pT)$, which form the \textquotedblleft ladder of amplitudes\textquotedblright  of harmonic oscillations,
		\begin{eqnarray}\label{eq:A.8}
		X_{(m_p)}[(m_p+1)T]={{\cal E}_0q\over M}\int_{m_pT}^{(m_p+1)T} K[(m_p+1)T-t_*] \sum_{m=0}^{m_p(t_*)}(-1)^m\epsilon_m\theta(t_*-mT) E^\prime_t(t_*-mT) dt_* ~~~~~~~~~~~~~~~~~~~~\nonumber\\
		+e^{-\Gamma_0T}\bigg\{[\cos\omega_0T+{\Gamma_0\over\omega_0}\sin\omega_0 T]\cdot {X}_{(m_p)}(m_pT)
		+\sin\omega_0 T\cdot {\dot{X}_{(m_p)}(m_pT)\over \omega_0}\bigg\} ,\nonumber~~~~~~\\
		{\dot{X}_{(m_p)}[(m_p+1)T]\over \omega_0}={{\cal E}_0q\over M} \int_{m_pT}^{(m_p+1)T} {\dot{K}[(m_p+1)T-t_*]\over \omega_0}\sum_{m=0}^{m_p(t_*)}(-1)^m\epsilon_m\theta(t_*-mT) E^\prime_t(t_*-mT)dt_* ~~~~~~~~~~~~~~~~~~~\\
		+ e^{-\Gamma_0 T}\bigg\{-(1+{\Gamma_0^2\over\omega_0^2} ) \sin\omega_0T\cdot {X}_{(m_p)}(m_pT)
		+  [\cos\omega_0T-{\Gamma_0\over\omega_0}\sin\omega_0T] \cdot {\dot{X}_{(m_p)}(m_pT) \over \omega_0}\bigg\}. \nonumber
		\end{eqnarray}
		As expected, the amplitude of the ladder decreases with a greater duration $T$ of its steps, which is an obvious effect of $\Gamma_0\neq 0$.
		We remind the reader that every term of the sequence $X_{(m_p)}$, $m_p=1,2,...$,  implicitly bears  the factor of ${{\cal E}_0q/ M}$, which initially appears in Eqs.(\ref{eq:A.3}.) for $m_p=0$, and is carried through by recursion (\ref{eq:A.8}).
			
	\section{Resonance factor ${\bm r(\omega)}$ in terms of ${\bm \zeta}$-variable}
	\label{app:appB}
	\renewcommand{\theequation}{B.\arabic{equation}}
	\setcounter{equation}{0}

	In terms of the variable $\zeta$, which is defined by the mapping (\ref{eq:E2.5}), $\uom=(\zeta+1/\zeta)/2$, the denominator of the  resonance factor $r(\omega)=[(\omega+i\Gamma_0)^2 -\omega_0^2]^{-1}$ becomes a fourth order polynomial with respect to $\zeta$. In what follows, $\uom_0=\omega_0/\Omega_e$ and $\uGa_0=\Gamma_0/\Omega_e$,
	\begin{eqnarray}\label{eq:B.1}
	r(\zeta) ={\zeta\over\Omega_e^2\uom_0}\bigg({1\over(\zeta-\zeta_1) (\zeta-\zeta_2)}
	-{1\over(\zeta-\zeta_3) (\zeta-\zeta_4)}\bigg)
	\end{eqnarray}
	with the roots $\zeta_{1,2}=(\uom_0-i\uGa_0)\pm i\sqrt{1-(\uom_0-i\uGa_0)^2}$ and $\zeta_{3,4}=-(\uom_0+i\uGa_0)\pm i\sqrt{1-(\uom_0+i\uGa_0)^2}$.   Since the function (\ref{eq:E5.5}) has no poles in the $\omega$-plane  and the radius of the contour $C_\zeta$ can be made arbitrary small,
	it is possible to expand the resonance factor $r(\zeta)$ in ascending powers of small $\zeta$. The algebra can be greatly simplified with an introduction of a complex angle, $\vartheta=\vartheta'+ i\vartheta''$, such that $\uom_0-i\uGa_0=\cos\vartheta$,  $\sqrt{1-(\uom_0-i\uGa_0)^2}=\sin\vartheta$ and, hence,
	$\zeta_{1}=e^{ i\vartheta} = e^{ i\vartheta'-\vartheta''} ,~~\zeta_2=1/\zeta_1,~~\zeta_{3}=-\zeta_1^\ast=
	-e^{- i\vartheta^\ast} ,~~\zeta_4= -\zeta_2^\ast=1/\zeta_3$. Then, Eq.~(\ref{eq:B.1}) can be rewritten as
	\begin{eqnarray}\label{eq:B.2}
	\Omega_e^2 \uom_0 r(\zeta)={\zeta \over (1-\zeta e^{i\vartheta}) (1-\zeta e^{-i\vartheta} )}-
	{\zeta \over (1+\zeta e^{i\vartheta^*}) (1+\zeta e^{-i\vartheta^*} )}=
	\sum_{k=2}^{\infty} \bigg[ {\sin k\vartheta\over\sin\vartheta} +(-1)^k{\sin k \vartheta^\ast\over\sin\vartheta^\ast} \bigg] \zeta^k\nonumber\\
	=\sum_{l=1}^{\infty}\bigg[2\;{\rm Re}\bigg({\sin2l\vartheta\over\sin\vartheta}\bigg)\zeta^{2l}+ 2i\;{\rm Im} \bigg({\sin(2l+1)\vartheta\over\sin\vartheta}\bigg)\zeta^{2l+1}\bigg]~,~~~~~~~~~~~~~~~~~~~~~~~~~~~~~~~~~~~~
	\end{eqnarray}
	where we  have noticed that the terms with $k=0$ and $k=1$ or, equivalently, with $l=0$ are zero. The Taylor series for the $r(\zeta)$ begins with term $\propto \zeta^2$.
	
	\section{Calculation of some integrals.}
	\label{app:appC}
	\renewcommand{\theequation}{C.\arabic{equation}}
	\setcounter{equation}{0}
	
   The double integral (\ref{eq:E5.10}) for $ {\cal E}_a$, is symmetric with respect to interchange $\omega\leftrightarrow\nu$. It is sufficient to consider only one of the two terms in the numerator,
	\begin{eqnarray}\label{eq:C.1}
	{{\sf d}{\cal E}_a(\tau,z<0)\over{\sf d}(\Omega_e\tz_0)}={2\Omega^2_q {\cal E}_0 \over 2\pi i\Omega_e}
	\sum_{m=0}^{m_p}\epsilon_m (-1)^m \bigg({-i\over 4\pi }\bigg)\oint_{C_\omega^-}{d\omega\over\omega} \mathfrak{T}[n_e(\omega)]e^{-i[\omega-\omega n_e(\omega)]\tz_0}	
	e^{-i\omega(\tau-|\tz| -2\tz_0-m T)}~~~~~~~\nonumber\\
	\times 	\oint_{C_\nu^-}{d\nu\over \nu} \mathfrak{T}[n_e(\nu)]{e^{-i[\nu-\nu n_e(\nu)]\tz_0}	\over \nu-\omega}.~~~~~~~~
	\end{eqnarray}
   The double integrals (\ref{eq:E5.11}) encountered for $ {\cal E}_b$, differ slightly,
	\begin{eqnarray}\label{eq:C.2}
	I_1=	\oint_{C_\omega^-}{d\omega\over\omega} \mathfrak{T}[n_e(\omega)] r(\omega)[2i\Gamma_0\omega-\omega_m^2]~e^{-i[\omega-\omega n_e(\omega)]\tz_0}
	e^{-i\omega (\tau-|\tz| -2\tz_0-m T)}
	\oint_{C_\nu^-}{d\nu\over \nu} \mathfrak{T}[n_e(\nu)]{e^{-i[\nu-\nu n_e(\nu)]\tz_0} \over \nu-\omega} ,\\
	\label{eq:C.3}	
	I_2=	\oint_{C_\nu^-}{d\nu\over \nu} \mathfrak{T}[n_e(\nu)]e^{-i[\nu-\nu n_e(\nu)]\tz_0}	
	e^{-i\nu(\tau-|\tz| -2\tz_0-m T)} 	
	\oint_{C_\omega^-}{d\omega\over\omega} \mathfrak{T}[n_e(\omega)] r(\omega)[2i\Gamma_0\omega-\omega_m^2]~{e^{-i[\omega-\omega n_e(\omega)]\tz_0} \over \nu-\omega} .
	\end{eqnarray}
	We use transformation (\ref{eq:E2.5}) to trade  $\nu$ in Eqs.(\ref{eq:C.1}) and (\ref{eq:C.2}) for a new variable $\zeta$.
	This leads to the integral over a circle of an arbitrary small radius around the origin,
	\begin{eqnarray}\label{eq:C.4}
	\oint_{C_\nu^-}{d\nu\over \nu} \mathfrak{T}[n_e(\nu)]{e^{-i[\nu-\nu n_e(\nu)]\tz_0} \over \nu-\omega} ~\to ~
	\oint^{(0+)} d\zeta~ {1-\zeta^2 \over \zeta^2-2\uom\zeta+1}~e^{-i\Omega_e\tz_0\zeta}=0.
	\end{eqnarray}
	This integral equals zero just because its integrand is a regular function inside the contour of integration.
	
	The integral in (\ref{eq:C.3}) differs from those in (\ref{eq:C.1}) and (\ref{eq:C.2}) by an additional factor $r(\omega)[2i\Gamma_0\omega-\omega_m^2]$ in the integrand. According to (\ref{eq:B.2}),  the Taylor expansion of the resonant factor $r(\omega)$ (in terms of variable $\zeta$) begins with $\zeta^2$, while $\omega\sim \zeta+\zeta^{-1}$, so that the Taylor expansion of the extra factor begins with $\zeta^1$. Hence, this integral is also zero.
	
	Integration (\ref{eq:E5.14}) is straightforward. Multiplying the result by external factor  $e^{-i\nu(\tau-|\tz|)}e^{+i[\nu +\nu n_e(\nu)]\tz_0}$ from Eq.(\ref{eq:E5.13}), we obtain,
\begin{eqnarray}\label{eq:C.5}
	e^{-i[\nu-\nu n_e(\nu)]\tz_0}\bigg[(C_1-iC_2)  {e^{i(\omega_0+i\Gamma_0)\tau_*}-	e^{-i\nu\tau_*}\over i(\nu+\omega_0+i\Gamma_0)}
	+(C_1+iC_2)  {e^{i(-\omega_0+i\Gamma_0)\tau_*}- e^{-i\nu\tau_*}\over i(\nu-\omega_0+i\Gamma_0)} \bigg],
\end{eqnarray}
where $\tau_*=t^*_{max}-t^*_{min}=\tau-|\tz|-2\tz_0-m_pT$ is the \textquotedblleft full time of radiation\textquotedblright. 	This function has no poles in the complex $\nu$-plane.
	
	\section{The source functions.}
	\label{app:appD}
	\renewcommand{\theequation}{D.\arabic{equation}}
	\setcounter{equation}{0}
	
	The results of calculations of Sec.\ref{sec:Sec6} are expressed via  four functions $s_j(\lambda,\xi)$, $j=1,2,3,4$. The functions  $s_j(\lambda,\xi)$  are {\em defined} as the sums of the following series,
	\begin{eqnarray}\label{eq:D.1}
		s_1(\lambda,\xi)=\sum_{l=1}^{\infty}(-1)^l {\rm Re}\bigg[{\sin2l\vartheta\over\sin\vartheta} \bigg]
		~[\xi^{2l}J_{2l}(\Omega_e\lambda)+\xi^{2l+2}J_{2l+2}(\Omega_e\lambda)],~~~~~~~~~~~~~~\nonumber\\   s_2(\lambda,\xi)=\sum_{l=1}^{\infty}(-1)^l  {\rm Re} \bigg[{\sin2l\vartheta\over\sin\vartheta} \bigg]
		~[\xi^{2l-1}J_{2l-1}(\Omega_e\lambda)-\xi^{2l+3}J_{2l+3}(\Omega_e\lambda)],~~~~~~~~\nonumber\\  s_3(\lambda,\xi)=\sum_{l=1}^{\infty}(-1)^l   {\rm Im}\bigg[ {\sin(2l+1)\vartheta\over\sin \vartheta}\bigg]  [\xi^{2l+1}J_{2l+1}(\Omega_e\lambda)+\xi^{2l+3}J_{2l+3}(\Omega_e\lambda)],~~\\
		s_4(\lambda,\xi)=\sum_{l=1}^{\infty}(-1)^l {\rm Im}\bigg[ {\sin(2l+1)\vartheta\over\sin \vartheta}\bigg]  [\xi^{2l}J_{2l}(\Omega_e\lambda)-\xi^{2l+4}J_{2l+4}(\Omega_e\lambda)].~~~~~~~~\nonumber
	\end{eqnarray}
	These sums  are intimately connected with the  Lommel functions $U_\nu(w,z)$ of two variables~\cite{Watson},
	\begin{eqnarray}\label{eq:D.2}
		W_\nu(w,z)=\sum_{l=0}^{\infty}w^{2l}J_{2l+\nu}(z) \equiv (iw)^{-\nu}U_\nu(iwz,z).
	\end{eqnarray}	

	Consider the calculation of $s_1$, as an example. Trading the original $\vartheta$ for $\pi/2+\delta$ (so that $\sin\delta =-\uom_0+i\uGa_0$) and exercising simple algebra, we arrive at,
	\begin{eqnarray}\label{eq:D.3}
		s_1(\lambda,\xi)=\sum_{l=1}^{\infty}{\rm Re}\bigg[{\sin2l\delta\over\cos\delta} \bigg]
		~[\xi^{2l}J_{2l}+\xi^{2l+2}J_{2l+2}]=\sum_{l'=0}^{\infty}
		{\rm Re}\bigg[{\sin(2l'+2)\delta +\sin2l'\delta\over\cos\delta}  \bigg]\xi^{2l'+2}J_{2l'+2}~~~~~~~~~~~~\nonumber\\
		=2\xi^2{\rm Re}\bigg[\sum_{l'=0}^{\infty}\sin[(2l'+1)\delta]\xi^{2l'}J_{2l'+2}\bigg]=
		2\xi^2{\rm Re}\bigg\{ {1\over 2i}\sum_{l'=0}^{\infty} \bigg[e^{i\delta}(e^{i\delta}\xi)^{2l'}-
		e^{-i\delta}(e^{-i\delta}\xi)^{2l'}\bigg]  J_{2l'+2}\bigg\},
	\end{eqnarray}
	where the argument $\Omega_e\lambda$ of the Bessel functions is omitted.
	Finally, referring to Eqs. (\ref{eq:D.2}),
	\begin{eqnarray}\label{eq:D.4}
		s_1(\lambda,\xi)= -\xi^2{\rm Re}\bigg\{ i\bigg[e^{i\delta}W_2(e^{i\delta}\xi,\lambda)
		-e^{-i\delta}W_2(e^{-i\delta}\xi,\lambda)\bigg]\bigg\}.
	\end{eqnarray}
	In the same way, rearranging the first sum in $s_2$ as $l\to l'+1$  and the second sum as   $l\to l'-1$ and subtracting the extra terms yields,
	\begin{eqnarray}\label{eq:D.5}
		s_2(\lambda,\xi)=2\uom_0\xi J_1(\Omega_e\lambda) +4{\rm Re}\bigg[\sin\delta \sum_{l'=0}^{\infty}\cos[2l'\delta]\xi^{2l'+1}J_{2l'+1}\bigg].~
	\end{eqnarray}
	which can be written as,
	\begin{eqnarray}\label{eq:D.6}
		s_2(\lambda,\xi) =+2\uom_0\xi J_1(\Omega_e\lambda)+2\xi {\rm Re}\big\{ \sin\delta \big[W_1(e^{i\delta}\xi,\lambda)
		+W_1(e^{-i\delta}\xi,\lambda)\big]\big\}.~~~~~~~~~~~~
	\end{eqnarray}
	The remaining two functions,  $s_3(\lambda,\xi)$ and  $s_4(\lambda,\xi)$  are transformed into
	\begin{eqnarray}\label{eq:D.7}
		s_3(\lambda,\xi)= 2 \sum_{l'=1}^{\infty}(-1)^{l'} {\rm Im}[\cos 2l'\vartheta ] \xi^{2l'+1}J_{2l'+1}
		=2\xi \sum_{l'=0}^{\infty}{\rm Im}[\cos 2l'\delta ] \xi^{2l'}J_{2l'+1}
		=\xi {\rm Im} \big[W_1(e^{i\delta}\xi,\lambda) +W_1(e^{-i\delta}\xi,\lambda)\big]~~~~~
	\end{eqnarray}
	and
	\begin{eqnarray}\label{eq:D.8}
		s_4(\lambda,\xi)\!= -4 {\rm Im}\bigg[\cos\vartheta \sum_{l'=0}^{\infty}(-1)^{l'} [\cos (2l'+1)\vartheta ]
		\xi^{2l'+2}J_{2l'+2}\bigg]=-4\xi ^2 {\rm Im}\bigg[\sin\delta\sum_{l'=0}^{\infty}\sin[(2l'+1)\delta] \xi^{2l'}J_{2l'+2}\bigg]  \nonumber\\	
		=\! 2\xi^2{\rm Im}\big\{ i\sin\delta\big[e^{i\delta}W_2(e^{i\delta}\xi,\lambda)
		\! -e^{-i\delta}W_2(e^{-i\delta}\xi,\lambda)\big]\big\}\!.~~~~~~~~~~~~~~~~~~~~~~~~
	\end{eqnarray}
	
	The functions  $s_j(\Lambda,\Xi)$ show up in  Eq.(\ref{eq:E5.19}) for the backward dipole radiation with the following arguments, $\Lambda^2=\Omega_e^2[(\tau-|\tz|-\tz_0-m_pT)^2-\tz_0^2]$, ~~
	$\Xi^2~=~(\tau-|\tz|-2\tz_0-m_pT)/(\tau-|\tz|-m_pT)$~ and
	$\Lambda\Xi =\Omega_e(\tau-|\tz|-2\tz_0-m_pT)$. \\
	In this study, the computational problems are somewhat alleviated by the fact, that we are interested in  the functions $s_j(\Lambda,\Xi)$ only at relatively small values of $\Lambda\Xi$.
	
	The Lommel functions of two variables, despite being named long ago, are not studied as exhaustively as, e.g. Bessel functions, and there are no tables (at least for the complex-valued variables) that could have been used for the numerical calculations. {\em A priori}, two practical methods seem obvious. One is to cut off the number of terms in the series (\ref{eq:D.2}). Another one is to use the integral representation for the Lommel functions (see Eq.\S 16.53(1) in Ref.\cite{Watson})
	\begin{eqnarray}\label{eq:D.9}
		W_\nu(\xi,\lambda)=\sum_{l=0}^{\infty}\xi^{2l}J_{2l+\nu}(\Omega_e\lambda)
		=\Omega_e\lambda\int_0^1 J_{\nu-1}(\Omega_e\lambda y) \cosh\big[{\Omega_e\lambda\xi\over 2} (1-y^2)\big]y^\nu dy~,~~ {\rm Re} (\nu)>0~,
	\end{eqnarray}
	In application to our problem, the major challenge in computing of these functions stems from the fact, that the Langmuir frequency, $\Omega_e$,  is very high ($\Omega_e/2\pi\sim 10^{15}$ Hz), so that
	the Bessel functions rapidly oscillate. Furthermore, in the integrand of (\ref{eq:D.9}), the amplitude of these oscillations grows exponentially when $y\to 0$ . Therefore, it is difficult to estimate the accuracy of the possible approximations. Here, we attempt to combine these methods.
		It is straightforward to check the following recursion formula,
	\begin{eqnarray}\label{eq:D.10}
	W_\nu(\xi,\lambda)=J_{\nu}(\Omega_e\lambda) +\xi^2W_{\nu+2}(\xi,\lambda).
	\end{eqnarray}
	By iterating the recursion formula (\ref{eq:D.10})  $N$ times  and applying  (\ref{eq:D.9}) to the $N+1$-st term,
	one readily obtains,
	\begin{eqnarray}\label{eq:D.11}
		W_\nu(\xi,\lambda)=
		\sum_{l=0}^{N-1}\xi^{2l}J_{2l+\nu}(\Omega_e\lambda)+\xi^{2N}\Omega_e\lambda\int_0^1 J_{\nu+2N-1}(\Omega_e\lambda y) \cosh\big[{\Omega_e\lambda\xi\over 2}(1-y^2)\big]y^{2N+\nu} dy.
	\end{eqnarray}
	When $N$ is sufficiently large, the exponential growth of hyperbolic cosine at $y\rightarrow 0$ and rapid oscillations of the Bessel function at $y\rightarrow 1$ in the integrand of $W_{\nu+2N}$ given by (\ref{eq:D.9}) become suppressed.
		
	In order to estimate an optimal value of the separation parameter $N$, let us notice that the lowest zero $j_\mu$ of the $J_\mu(z)$ and the first maximum, $j^\prime_\mu$ of the $J^\prime_\mu(z)$,   are greater than $\mu$ (see. \cite{Watson}, \S 15.3(1)), $$j_\mu>\mu,~j^\prime_\mu>\mu.$$
	For functions of large order, a simple estimate of the smallest zero and the smallest maximum is as follows (\cite{Watson}, \S15.83),
	\begin{eqnarray}\label{eq:D.12}
		j_\mu=\mu+1.855757\times\mu^{1/3} + O(\mu^{-1/3}), ~~~~	j^\prime_\mu=\mu+0.808618\times\mu^{1/3} + O(\mu^{-1/3}).
	\end{eqnarray}
In order that there are no zeros of the Bessel function within the interval of integration over $y$ in (\ref{eq:D.11}),  it is necessary that the argument of the Bessel functions does not exceed its smallest zero or its smallest maximum, i.e, $\Omega_e\lambda y <\Omega_e\lambda \leq j_{2N+\nu-1}$ or $\Omega_e\lambda \leq j'_{2N+\nu-1}$. Hence, the integral accommodates that part of the sum, where order of the Bessel functions exceeds its argument.
The simplest estimate of $N=N(m,T)$ is given by the equations,
	$$\Omega_e\lambda\leq j_{2N+\nu-1}\sim 2N, ~~~\Omega_e\lambda\leq j'_{2N+\nu-1}\sim 2N.$$
Numerical calculation show, that for sufficiently large upper limit $N(T)$ of the sum over $l$, the integral in  Eq.(\ref{eq:D.11}) is small.

A few remarks regarding asymptotic behavior of the source functions, which clarify the origin of their behavior, observed in Figs.\ref{Fig:fig8} and \ref{Fig:fig9}, which are based on numerical calculations and presented in Sec.\ref{subsec:Sec6A}, are in order. The period of plasma oscillation is $\tau_e=2\pi/\Omega_e\sim 5\cdot10^{-16} - 10^{-15}$sec and the corresponding unit of length is $\lambda_e=c\tau_e\sim 10^{-5} - 10^{-4}{\rm cm}\approx 10^{3} - 10^{4}$\AA. 	By  nature of our problem, we are interested in the time interval  $T_0=2\pi\omega_0\sim 10^3\tau_e$, so that $\Xi^2=1-2\tz_0/\tau_m$ is very close to the constant value of $1$, while $\Lambda\approx \tau_m=\tau-|\tz|-m_pT$. Let us consider the limit of $\Xi=1$ as the zero-order approximation when $\tau_m\gg z_0$. Curiously enough, it coincides with the exact solution with $z_0=0$, which corresponds to the location of the radiating dipole on the interface between vacuum and a medium.  Then the functions $W_\nu(e^{i\delta},\Lambda)$, which,  according to (\ref{eq:D.2}),  $(ie^{i\delta})^\nu W_\nu(e^{i\delta},\Lambda)= U_\nu(ie^{i\delta}\Lambda,\Lambda),$ are the Lommel's functions $y=U_\nu(c\Lambda,\Lambda)$ of two variables with $w=c\Lambda$, where $c$ is constant. In our case, $c=ie^{i\delta}$, so that  $(c+c^{-1})^2=4\sin^2\delta$ and $y=(ie^{i\delta})^\nu  W_\nu(e^{i\delta},\Lambda)$~
These functions are particular integrals of the equation  for $y=U_\nu(c\Lambda,\Lambda)$ (\cite{Watson}, \S16.52 (7)). The function $W_\nu(e^{i\delta},\Lambda)$ satisfy the following equation,
  \begin{eqnarray}\label{eq:D.13}
4\big\{ {d^2 W_\nu(e^{i\delta},\Lambda)/ d  \Lambda^2}+\sin^2\delta W_\nu(e^{i\delta},\Lambda)\big\}= J_{\nu-2}(\Omega_e\Lambda)-e^{-2i\delta}J_\nu(\Omega_e\Lambda),
 \end{eqnarray}
 which obviously has, among others, the periodic solutions like $\cos(\Omega_e\Lambda\sin\delta)=\cos(\omega_0\Lambda)$.

When $\tau\gg z_0$ it is instructive to present the functions $s_j(\Lambda, 1)$ in a somewhat different form,
	\begin{eqnarray}\label{eq:D.14}
	s_1(\lambda,1)=2{\rm Re}\bigg\{\sum_{l'=0}^{\infty}\sin[(2l'+1)\delta]J_{2l'+2}(\Omega_e\Lambda)\bigg\}
	=2{\rm Re}\bigg\{\sum_{l'=1}^{\infty}\sin[(2l'-1)\delta]J_{2l'}(\Omega_e\Lambda)\bigg\}~~~~\\
	= 2{\rm Re}\bigg\{-\sin\delta \sum_{l=1}^{\infty}\cos2l\delta~J_{2l}(\Omega_e\Lambda)
	+\cos\delta \sum_{l=0}^{\infty}\sin[(2l+2)\delta] ~J_{2l+2}(\Omega_e\Lambda)\bigg\} ~~~~~~~~~~,\nonumber
	\end{eqnarray}
    The first sum in the last equation is well known to be
	$ 2\sum_{l=1}^{\infty}\cos2l\delta J_{2l}(\Omega_e\lambda)=\cos(\Omega_e\lambda\sin\delta)- J_0(\Omega_e\lambda)$, while the second sum differs from the original one by replacement $\sin[(2l+1)\delta]\to \sin[(2l+2)\delta]$. The function $s_1(\Lambda,1)$ can be cast as	
	\begin{eqnarray}\label{eq:D.15}
	s_1(\Lambda,1)	= {\rm Re}\big\{\sin\delta[J_0(\Omega_e\Lambda)-\cos(\Omega_e\Lambda\sin\delta)] \big\}
-{\rm Re}\big\{ i\cos\delta\big[e^{2i\delta}W_2(e^{i\delta},\Lambda)
	-e^{-2i\delta}W_2(e^{-i\delta},\Lambda)\big]\big\},
	\end{eqnarray}
	where $\Lambda\approx\tau_m$ and $\Omega_e\sin\delta\approx\omega_0$. In agreement with numerical calculations, the source function $s_1(\Lambda,1)$ contains an observed sum of slow harmonic and a precursor of the dipole radiation.
	
	In the same way, since in Eq.(\ref{eq:D.5})  $\cos 2l\delta=\cos\delta \cos(2l+1)\delta +\sin\delta \sin(2l+1)\delta$, the source function $s_2(\Lambda,1)$ can be written down as
\begin{eqnarray}\label{eq:D.16}
	s_2(\Lambda,1) =2\uom_0J_1(\Omega_e\Lambda)+{\rm Re}\bigg \{4 \sin^2\delta \sum_{l=0}^{\infty} \sin[(2l+1)\delta]J_{2l+1}(\Omega_e\Lambda)
+2 \sin 2\delta\sum_{l=0}^{\infty} \cos[(2l+1)\delta]J_{2l+1}(\Omega_e\Lambda)\bigg\}~ \\
=-2\uom_0J_1(\Omega_e\Lambda)+2{\rm Re}\{\sin^2\delta\sin(\Omega_e\Lambda\sin\delta)\}
+2 {\rm Re}\big\{ \sin 2\delta\big[e^{i\delta}W_1(e^{i\delta},\Lambda)
+e^{-i\delta}W_1(e^{-i\delta},\Lambda)\big]\big\}.~~~~~~~\nonumber
\end{eqnarray}
where we employed another well-known result,	$ 2\sum_{l=0}^{\infty}\sin[(2l+1)\delta]J_{2l+1}(\Omega_e\Lambda) =\sin(\Omega_e\Lambda\sin\delta)\approx\sin(\omega_0\Lambda) $.

Using the same transformations, it is straightforward to obtain,
\begin{eqnarray}\label{eq:D.17}
	s_3(\Lambda,1)= 2\sum_{l=0}^{\infty}{\rm Im}[\cos 2l\delta ]J_{2l+1}(\Omega_e\Lambda)
=2{\rm Im}\bigg\{\sum_{l=0}^{\infty} [\cos\delta\cos (2l+1)\delta
+\sin\delta\sin(2l+1)\delta] J_{2l+1}(\Omega_e\Lambda)\bigg\}~~~~~~~~~~~\\
=   {\rm Im} \{  \sin\delta \sin(\Omega_e\Lambda\sin\delta)\}
+ {\rm Im}\bigg\{\cos\delta\big[e^{i\delta}W_1(e^{i\delta},\Lambda) +e^{-i\delta}W_1(e^{-i\delta},\Lambda)\big] \bigg\} ~~~~~~~~~~~~~~~~~~~~~~~~~~~\nonumber
\end{eqnarray}
and
\begin{eqnarray}\label{eq:D.18}
s_4(\Lambda,1) =-4{\rm Im}\bigg\{\sin\delta\sum_{l'=0}^{\infty}\sin[(2l'+1)\delta] J_{2l'+2}(\Omega_e\Lambda) \bigg\}	
= 4  {\rm Im}\bigg\{\sin\delta\sum_{l=0}^{\infty}[\sin\delta\cos 2l\delta
 -\cos\delta \sin 2l\delta ]J_{2l}(\Omega_e\Lambda) \bigg\}  ~\nonumber\\
=2{\rm Im}\big\{ \sin^2\delta [\cos(\Omega_e\Lambda\sin\delta)-J_0(\Omega_e\Lambda)]\big\}
+ 2{\rm Im}\big\{ i\sin2\delta\big[e^{2i\delta}W_2(e^{i\delta},\Lambda)
-e^{-2i\delta}W_2(e^{-i\delta},\Lambda)\big]\big\}.~~~~~~~~~~~~
\end{eqnarray}

\end{document}